\documentclass[pdflatex,sn-basic]{sn-jnl}

\usepackage{graphicx}%
\usepackage{multirow}%
\usepackage{amsmath,amssymb,amsfonts}%
\usepackage{amsthm}%
\usepackage{mathrsfs}%
\usepackage[title]{appendix}%
\usepackage{xcolor}%
\usepackage{textcomp}%
\usepackage{manyfoot}%
\usepackage{booktabs}%
\usepackage{algorithm}%
\usepackage{algorithmicx}%
\usepackage{algpseudocode}%
\usepackage{listings}%
\usepackage{cleveref}

\theoremstyle{thmstyleone}%
\theoremstyle{thmstyletwo}%

\theoremstyle{thmstylethree}%

\newcommand{\sech}[0]{\text{sech}}

\crefname{figure}{Figure}{Figures}
\Crefname{figure}{Figure}{Figures}
\crefname{table}{Table}{Tables}
\Crefname{table}{Table}{Tables}
\crefname{section}{Section}{Sections}
\Crefname{section}{Section}{Sections}
\crefname{equation}{}{}
\Crefname{equation}{Equation}{Equations}
\crefname{appendix}{Appendix}{Appendices}
\Crefname{appendix}{Appendix}{Appendices}

\makeatletter
\AddToHook{cmd/appendix/before}{\def\cref@section@alias{appendix}}
\makeatother

\begin{document}

\graphicspath{{Figures}}

\title[F-actin and cell motility]{How negative feedback from filamentous actin affects cell shapes and motility}


\author*[1]{\fnm{Jack M.} \sur{Hughes}}\email{jack.hughes@mail.concordia.ca}

\author[2]{\fnm{Jupiter} \sur{Algorta}}\email{jupitera@math.ubc.ca}

\author[2]{\fnm{Leah} \sur{Edelstein-Keshet}}\email{keshet@math.ubc.ca}

\affil*[1]{\orgdiv{Department of Mathematics and Statistics}, \orgname{Concordia University}, \orgaddress{\street{1455 De Maisonneuve Blvd. W.}, \city{Montr\'eal}, \postcode{H3G 1M8}, \state{QC}, \country{Canada}}}

\affil[2]{\orgdiv{Department of Mathematics}, \orgname{University of British Columbia}, \orgaddress{\street{1984 Mathematics Rd.}, \city{Vancouver}, \postcode{V6T 1Z2}, \state{BC}, \country{Canada}}}

\abstract{The crawling motility of many eukaryotic cells is driven by filamentous actin (F-actin), and regulated by a network of signaling proteins and lipids (including small GTPases). The tangle of positive and negative feedback loops gives rise to various experimentally observed dynamic patterns (\emph{actin waves}). Here we consider a recent prototypical model for actin waves in which F-actin exerts negative feedback onto a GTPase. Guided by previous numerical PDE bifurcation analysis, we explore cell shapes and motility associated with polar, oscillatory, and traveling wave solutions of a mass-conserved partial differential equation (PDE) model. We use Morpheus (cellular Potts) simulations to investigate the implications of such regimes of behavior on the shapes and motion of cells, and on transitions between modes of behavior. The model demonstrates various cell states, including resting (spatially uniform GTPase activity), polar cells (static zones of GTPase activity), and traveling waves along the cell edge. In some parameter regimes, such states can coexist, so that cells can transition from one behavior to another in response to noisy stimuli.
}




\maketitle

\textbf{Acknowledgments} We are grateful to Arik Yochelis for helpful discussions and comments. This work was funded by a Natural Sciences and Engineering Research Council of Canada (NSERC) CGS-D Scholarship, and Centre de Recherches Math\'{e}matiques and the Institut des Sciences Math\'{e}matiques postdoctoral fellowship awarded to JMH and an NSERC Discovery Grant to LEK.

\section{Introduction}

Understanding eukaryotic cell motility continues to be one of the grand challenges of modern cell biology, engendering both experimental and theoretical research. Experimentally, many complex cell motility modes have been observed in a wide variety of cell types~\citep{barnhart2017adhesion,merino2022review,Moldenhawer2022Switching,town2023Rac}. Commonly, cells experience directed cell motion, which is key in many biological processes including cancer metastasis~\citep{spano2012cancer} and wound healing~\citep{weiner1999spatial,wong2006neutrophil}. Other modes of motility can also occur such as random amoeboid-type motion, and cell turning and ruffling. Among the fundamental questions to be addressed, we ask how cell motility is regulated and what causes cells to change motility modes.

While new mechanisms powering cell motility are continually being discovered, the leading role of filamantous actin (F-actin) is well-recognized in basic directed migration of neutrophils, keratocytes, social amoeba (\emph{Dictyostelum discoideum}) and other cell types. F-actin is a key player in cell motility because the growing ends of these filaments, at which actin monomers assemble, accumulate near the front edge of a cell and cause the edge to protrude outwards. However, it is not always clear what interactions between F-actin and its regulators are at the core of the complex motility machinery. Our interest here is in accounting for basic cell motility patterns such as directed motion, turning, ruffling, and random motility. We consider a small subset of the central signaling network, consisting of a small GTPase (such as Rho, Rac, or Cdc42) (\cref{fig:SchematicJHBio}a) and the interaction with F-actin. 

GTPases act as molecular switches, with the active form bound to the membrane and the inactive in the cell interior (cytosol). The activation and inactivation of these GTPases are on the scale of seconds/minutes, while any loss or production is on a much longer scale~\citep{Machacek2009GTPase}. This justifies the common modeling assumption that the total quantity of a given GTPase is conserved over the timescale of interest for motile cells (minutes). The conversion from inactive to active form is governed by GEFs (guanine nucleotide exchange factors) and the reverse by GAPs (GTPase-activating proteins). The patterns induced by these Rho-GTPases dictate the distribution of F-actin.

Recent experimental evidence shows that F-actin is not only governed by Rho GTPases but, in certain cells, it can also affect the activity of the GTPase. Feedback between GTPases and F-actin results in a variety of experimentally observable dynamic patterns, commonly denoted \emph{actin waves}~\citep{Allard2013,Arai2010,barnhart2017adhesion,Bement2015,bernitt2017fronts,beta2017intracellular,dobereiner2006lateral,gerisch2009self,giannone2004periodic,Landino2021,Michaud2021,maxian2025actin,vicker_f_actin_2002,yang2019two}. For example, F-actin can promote GTPase inactivation by recruiting GAPs~\citep{bement2024patterning} (\cref{fig:SchematicJHBio}b). This negative feedback loop leads to many interesting actin wave structures and forms the basis of the actin waves modeling in this paper. Experimental work shows a rich structure of waves along the cell edge, including oscillatory waves~\citep{giannone2004periodic}, traveling waves along the cell front~\citep{barnhart2017adhesion} and waves on the cell membrane~\citep{yang2018rhythmicity,Michaud2022}. Thus, these wave-like dynamics appear in many different settings, although the proposed mechanisms governing these patterns vary. A key motivation of our work is to explore how these actin waves affect cell motility.

Models describing actin waves tend to be complex, often relying on numerical simulations alone to understand model behavior. Such studies favor model details over unraveling generic model structure. It can be challenging to fully understand the results. Using simulations alone can also miss important parameter regimes where there is coexistence of different behaviors or transitions from one behavior to another. In this work, we leverage a previous study that analyzed a simple reaction-diffusion (RD) model for actin waves \citep{hughes2024travelling}. There, numerical bifurcation analysis was used to identify a mathematical mechanism for coexistence of polar patterns (with a clear front and back) and multi-peak traveling waves suspected to lead to directed cell motion and cell ruffling, respectively. In particular, a parameter regime was identified where such patterns are stable and coexist. Other works in which numerical bifurcation analysis has been used to study actin waves models include \cite{bernitt2017fronts,Yochelis2022}. Here, our key motivation is to then determine what types of cell shapes and motility occur based on the structures identified by~\cite{hughes2024travelling}.

The intracellular dynamics of actin waves only provides part of the picture at the cell-scale. To understand how actin waves affect cell shape and motility, the dynamics must be coupled to a model for cell deformations. Common techniques for capturing deformable cell domains include Lagrangian marker point, level set methods~\citep{neilson2011cell}, phase field methods~\citep{shao2012motility,camley2017crawling,alonso2018modeling}, finite elements~\citep{cusseddu2019bulksurface,elliott2012motility}, and Metropolis-based methods including the cellular Potts model (CPM)~\citep{algorta2025investigating}. Additional references using such methods can be found in the reviews by~\cite{alert2020review,beta2023actin,Buttenschon2020,camley2017review,holmes2012shape,spatarelu2019review,sun2016review}. The goal of the present study is to provide a simple, reproducible connection between the reaction-diffusion dynamics and cell deformations so that cell-scale structures can be studied. Therefore, we use the CPM framework~\citep{graner1992CPM}, as this is arguably the simplest~\citep{TeamKeshet2024} and can be easily implemented using the open source software package Morpheus~\citep{starruss2014morpheus}. For the basics on the CPM framework, see~\cref{sec:sim motility}.

Many previous papers have described dynamic cell shapes using the CPM formalism, and many of those papers have included PDEs for the proteins that regulate F-actin and cell motion. Examples include \cite{maree2012cells,Liu2021,Rens_Edelstein-Keshet_2021}. These examples are more advanced in that they simulated PDEs inside fully 2D deforming cell domains. However, none of these works had accompanying PDE bifurcation analysis, nor did they attempt to map out the mechanisms for transitions between multiple observed cell behaviors. Also, since previous simulations were done with custom-made proprietary codes, reproducing previous results, or building upon them, remains challenging.

To address these issues, we use the open source simulation platform Morpheus, together with the numerical bifurcation analysis by \cite{hughes2024travelling}. To investigate cell shape and motility, we construct initial conditions and choose parameter values near key locations along the solution branches. We depict cell shapes using CPM simulations, where outward protrusion updates are favored in areas with high F-actin concentration. Our results demonstrate transitions between directed cell motion, cell turning, and ruffling based on the intracellular dynamics of actin waves.

\section{Model description and previous results}
\subsection{The signaling model} \label{sec:signal model}
\begin{figure}[!tp]
    \centering
    \includegraphics[width=\linewidth]{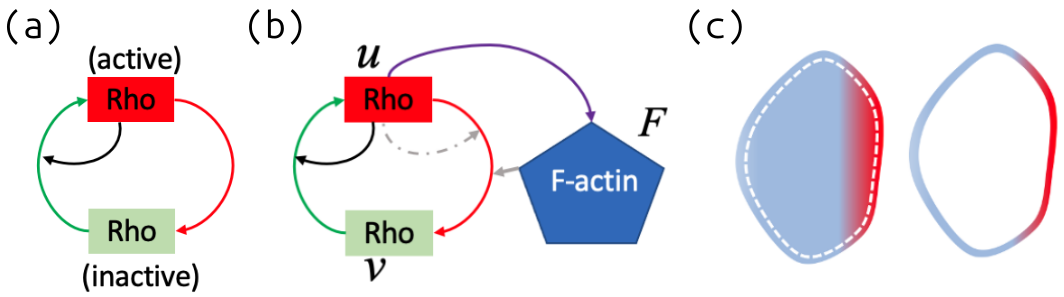}
    \caption{\textbf{Schematic diagram of models:} (a) Typical cycling between active (red) and inactive (green) Rho GTPase with positive feedback from the active form to its own activation. The model by \cite{Mori2008} for polarization (\emph{wave-pinning}, WP) was based on this circuit. (b) A downstream effector (such as F-actin) is promoted by the GTPase, and then exerts negative feedback (enhances the inactivation of Rho).
    Arrow colors represent GAP (red), GEF (green), downstream target (purple), positive (solid black) and negative (gray) feedback. The dot-dashed gray arrow in (b) was proposed as a simplification of the saturating kinetics used in \cite{holmes2012}. (c) The distribution of active GTPase on a thin sheet-like cell has been simulated elsewhere in 2D (e.g., in \cite{maree2012cells}). Here, our geometry is a periodic 1D domain along the cell edge (white dashed curve, and contour shown on the right).
      }
    \label{fig:SchematicJHBio}
\end{figure}

The model of interest, proposed in~\cite{hughes2024travelling}, consists of two underlying subsystems. The first is the main regulatory unit (e.g. Rho GTPase) whose active ($u$) and inactive ($v$) forms satisfy overall mass conservation (\cref{fig:SchematicJHBio}a). This subsystem, denoted \emph{wave-pinning} was studied by \cite{Mori2008} as a model for cell polarization: the PDE model supports a stable pattern with a zone of high Rho activity under appropriate conditions. In fully dimension-carrying form, the model is given by the set of PDEs
\begin{subequations}\label{eq:MoriModel}
    \begin{align} 
    \frac{\partial u}{\partial t}&= \left(\beta_0+\gamma_0\frac{u^2}{u_0^2+u^2}\right)v -\delta_0u + D_{u,0}\frac{\partial^2u}{\partial x^2},\label{eq:Mori u}\\
    \frac{\partial v}{\partial t}&= -\left(\beta_0+\gamma_0\frac{u^2}{u_0^2+u^2}\right)v+ \delta_0u+D_{v,0}\frac{\partial^2v}{\partial x^2},\label{eq:Mori v}
\end{align}
\end{subequations}
where $\beta_0$ is a basal rate of activation, $\delta_0$ is an inactivation rate, $\gamma_0$ is the rate of auto-activation (positive feedback of active GTPase to accelerate its activation rate), $u_0$ is the EC50 parameter, that is, the GTPase level at which the Hill function feedback term is at its half maximum, and $D_{u,0}<D_{v,0}$ are the rates of diffusion for the active and inactive GTPase. Here the model assumes positive feedback onto the GEFs. See Appendix A in \cite{jilkine2007mathematical}, which shows that negative feedback onto the GAPs is qualitatively the same. We use periodic boundary conditions on a spatial domain $(0,L)$ to represent the edge of a 2-dimensional cell shape, as shown in Figure~\ref{fig:SchematicJHBio}c.
In this geometry, a zone of active Rho corresponds to a peak or plateau of $u$.

This model conserves mass:
\begin{align} \label{eq:mass con}
G_T:=\frac{1}{L}\int_0^L\left[u(x,t)+v(x,t)\right]{\textrm d}x=\text{constant},
\end{align}
where $(0,L)$ is the spatial domain and $L$ denotes the perimeter of the cell. The mass is normalized by the domain length and thus $L\cdot G_T$ denotes the total, conserved amount of GTPase in the system. The advantage of starting with this form of the model~\cref{eq:MoriModel} is that we know the interpretation and units of all the basic parameters. Namely, $\beta_0, \gamma_0, \delta_0$ have units of 1/time, $u, v, u_0$ have units of GTPase concentration, and $D_{u,0}$ and $D_{v,0}$ have units of length$^2$/time. 

We rewrite the reaction kinetics by rescaling the Hill function
\begin{align*}
    f(u,v):=\left(\beta_0+\gamma_0\frac{u^2}{u_0^2+u^2}\right)v -\delta_0u=\left(\beta_0+\gamma_0\frac{(u/u_0)^2}{1+(u/u_0)^2}\right)v -\delta_0u,
\end{align*}
to have a dimensionless numerator and denominator. 
In the above original wave-pinning model, \eqref{eq:MoriModel}, unlimited growth of Rac is prevented by assuming that self-activation is limited (using a Hill-function). In the related model analysed by \cite{hughes2024travelling}, this assumption is relaxed by balancing auto-activation with an auto-inhibition terms. Indeed, \cite{hughes2024travelling},  proposed polynomial terms in place of the kinetics of~\eqref{eq:MoriModel}, simplifying the PDEs to  
\begin{subequations}\label{eq:MoriModel poly}
    \begin{align} 
    \frac{\partial u}{\partial t}&= \left(\beta_0 + \gamma_1 (u/u_0)^2 \right) v - \delta_0 (1+(u/u_0)^2) u + D_{u,0} \frac{\partial^2 u}{\partial x^2},\\
    \frac{\partial v}{\partial t}&= -\left(\beta_0 + \gamma_1 (u/u_0)^2 \right) v + \delta_0 (1+(u/u_0)^2) u + D_{v,0} \frac{\partial^2 v}{\partial x^2},
\end{align}
\end{subequations}
where $\gamma_1= (\beta_0 + \gamma_0)$.
The modified model \eqref{eq:MoriModel poly} preserves the same spatially uniform steady states as the original WP model. Further, it is more closely related to well-known prototypical models~\citep{jd2003mathematical,murray2007mathematical} analogous to the FitzHugh-Nagumo and Gray-Scott models.

Analogously to the derivation of an existing model for actin waves in  \cite{holmes2012}, we couple the GTPase variables $u,v$ to F-actin ($F$) to arrive at the dimensional PDEs
\begin{subequations}
    \begin{align}
        \frac{\partial u}{\partial t}&= \left(\beta_0 + \gamma_1 (u/u_0)^2 \right) v - \delta_0 (1+(u/u_0)^2 + \sigma F) u  + D_{u,0} \frac{\partial^2 u}{\partial x^2},\\
        \frac{\partial u}{\partial t}&= -\left(\beta_0 + \gamma_1 (u/u_0)^2 \right) v + \delta_0 (1+(u/u_0)^2 + \sigma F) u  + D_{v,0} \frac{\partial^2 u}{\partial x^2},\\
        \frac{\partial F}{\partial t}&= \theta(\rho_0+\rho_1u-F)+D_{F,0}\frac{\partial^2 F}{\partial x^2},
    \end{align}
\end{subequations}
where $\delta_0\sigma$ is the F-actin dependent GTPase inactivation rate (\cref{fig:SchematicJHBio}b), $\theta\rho_0,\theta\rho_1$ denote the basal and GTPase dependent F-actin assembly rates, and $\theta$ denotes the F-actin disassembly rate. The F-actin diffusion coefficient $D_{F,0}$ is assumed to be very small or negligible. To nondimensionalize, we define the dimensionless variables
\[
u^\ast:=\frac{u}{u_0},\quad v^\ast := \frac{v}{u_0}, \quad F^\ast:=\frac{F}{F_0}, \quad t^\ast:=\delta_0 t, \quad x^\ast:=\frac{x}{L},
\]
where $F_0$ is the scale for $F$ (not used in~\cite{hughes2024travelling}). Note that $u$ and $v$ have the same scale. Substituting into the model PDEs~\cref{eq:MoriModel poly}, simplifying and dropping the $\ast$'s yields our GTPase-actin model,
\begin{subequations}\label{eq:model}
    \begin{align} 
    \frac{\partial u}{\partial t}&= \left(b+ \gamma u^2 \right)  v -  (1+u^2 + s F) u  + D_u\frac{\partial^2 u}{\partial  x^2},\\
    \frac{\partial v}{\partial t}&= -\left(b+ \gamma u^2 \right)  v +  (1+u^2 + s F) u  + D_v\frac{\partial^2 v}{\partial  x^2},\\
    \frac{d F}{d t}&= \omega (p_0+p_1 u - F)+D_F\frac{\partial^2 F}{\partial x^2},
\end{align}
\end{subequations}
where
\begin{gather*}
b:=\frac{\beta_0}{\delta_0},\quad \gamma:=\frac{\gamma_1}{\delta_0}=\frac{\beta_0+\gamma_0}{\delta_0}, \quad s:=\sigma F_0, \quad \omega:=\frac{\theta}{\delta_0},\quad p_0:=\frac{\rho_0}{F_0},\\
p_1:=\frac{\rho_1u_0}{F_0},\quad \quad D_u:=\frac{D_{u,0}}{\delta_0L^2}, \quad D_v:=\frac{D_{v,0}}{\delta_0L^2}, \quad D_F:=\frac{D_{F,0}}{\delta_0L^2},
\end{gather*}
are dimensionless parameters with $F_0:=1$. (With a proper choice of $F_0$, an additional parameter could be scaled out.) In the rescaled variables, mass conservation becomes
\begin{align*}
    \int_0^1[u(x,t)+v(x,t)]{\textrm d}x=\frac{G_T}{u_0}=:M. 
\end{align*}
For mathematical convenience and to apply the results of~\cite{hughes2024travelling}, we rescale the domain length back to $L$. To retain our model structure~\eqref{eq:model}, we can then scale the non-dimensional diffusion coefficients, $D_i=D_{i,0}/(\delta_0L^2)$ for $i=\{u,v,F\}$ accordingly. Note that by choosing different kinetic rates as our time scale, $\delta_0$ could be replaced by the basal GTPase activation rate, $\beta_0$, or the self-amplifying GTPase activation rate, $\gamma_0$.

The basic interactions of the model \cref{eq:model} (\cref{fig:SchematicJHBio}b) were found experimentally in \cite{Bement2015, Goryachev2016,Landino2021,Michaud2021, swider_cell_2022}, and modeled in a similar spirit in these and companion papers.
In this system, the GTPase is RhoA, and F-actin its downstream effector. Rho activates its own GEF, Ect2. Rho promotes F-actin through the formin (mDia) and F-actin recruits the GAP RGA3/4 that inactivates RhoA.

While the model in the above papers differs from ours in various details, the basic features are quite similar. We can identify our parameter $b$ as a basal rate and $\gamma$ as the Ect2-dependent Rho activation rate (Rho self-activates via Ect2). The parameter $s$ is the RGA3/4-dependent Rho inactivation rate, which is assumed to be proportional to the level of F-actin. The total Rho (active plus inactive) is $M$. 

\begin{table}[htbp]
\footnotesize
\caption{\textbf{Model parameter values:} Dimensionless parameter values used in simulations} \label{tab:par valuesSims}
\centering
    \begin{tabular}{|c|l|c|} \hline
        \bf Parameter & \bf Definition & \bf Value\\ \hline
        $b$ & GTPase basal activation rate & varied\\
        $\gamma$ & GTPase feedback activation rate & 3.557\\
        $s$ & F-actin dependent inactivation rate& varied\\
        $\omega$ & F-actin time scale parameter& 0.6\\
        $p_0$ & F-actin basal growth rate & 0.8\\
        $p_1$ & GTPase dependent F-actin growth rate & 3.8\\
        $L$ & Cell perimeter & varied \\
        $D_u$ & Active GTPase diffusion rate & 0.1$/L^2$\\
        $D_v$ & Inactive GTPase diffusion rate & 1$/L^2$\\
        $D_F$ & F-actin diffusion rate & 0.001$/L^2$ \\ 
        $M$ & Average GTPase density & 2 \\ \hline
    \end{tabular}
\end{table}

\subsection{Summary of previous PDE bifurcation results}
Our GTPase-actin model~\cref{eq:model} has a rich bifurcation structure of both propagating and stationary patterns~\citep{hughes2024travelling}. Important solutions for this work are spatially periodic traveling waves with a fixed profile and speed, and stationary polar patterns with a plateau of high concentration flanked by plateaus of low concentration. The traveling waves emerge from many distinct bifurcations, including finite wavenumber oscillatory (Hopf) bifurcations of the homogeneous steady state (HSS) with an intrinsic wavelength~\citep{cross1993pattern}. The polar patterns emerge from saddle-nodes of stationary pulse solutions, and the stationary pulses emerge from long wavelength bifurcations of the HSS (see~\cite{hughes2024travelling} for more details). The signature bifurcations of the HSSs that may lead to stable traveling waves and polar patterns are thus finite wavenumber Hopf and long wavelength bifurcations, respectively. To identify a parameter regime where traveling waves and polar patterns coexist,~\cite{hughes2024travelling} varied $s$ and $b$, and found a codimension-2 bifurcation with simultaneous emergence of a finite wavenumber Hopf and long wavelength instability (see~\cref{app:codim2} for the parameter values). The coexistence of the emerging solutions from this codimension-2 bifurcation are then shown by computing a one-parameter bifurcation diagram in $s$. The traveling waves in this regime have a wavelength of $\lambda$. The polar patterns with a wavelength $\lambda$ are all unstable; however, stable polar patterns exist with wavelengths $m\lambda$ for $m=2,3,4,...$. Coexistence then requires traveling waves be periodically extended to fit into the same domain, indicating the emergence of ruffling cell motility. This coexistence regime indicates that cells can switch between directed motion and cell ruffling with sufficiently noisy stimuli. Following~\cite{hughes2024travelling}, we focus on the domain lengths $L=2\lambda,3\lambda$, as these produce coexistence regimes but have different transient dynamics that can lead to distinct cell motility modes.

When $L=3\lambda$, an additional behavior is observed where traveling waves with a wavelength $3\lambda/2$ emerge from a parity-breaking bifurcation of polar patterns instead of a finite wavenumber Hopf (see~\cref{sec:unsteady sims} for more details). These polar patterns also have a wavelength of $3\lambda/2$ suggesting a possible mechanism to transition between directed motion and cell turning with a time-varying $s$ that crosses the parity-breaking bifurcation. However, these polar and traveling wave patterns are unstable in this regime, and they do not seem to stabilize for $L=m\lambda$ with $m=1,2,3,...$. We show that for domain lengths between $L=\lambda$ and $L=2\lambda$, these solutions can be stable. We use $L=5$ to demonstrate that transitions between directed motion and cell turning can occur. A domain length of $L=5$ is chosen since the stability region of the polar patterns is too small for $L=3\lambda/2\approx4.6$.

Following~\cite{hughes2024travelling}, we choose $s$ and $b$ as our main and secondary parameter, respectively, and demonstrate how different domain lengths can affect cell shapes and motility. These parameters are chosen based on previous studies of similar models~\citep{holmes2012,Liu2021,mata2013model}. The results for $L=3\lambda,2\lambda,5$ are shown in \cref{sec:CPMResults,sec:smaller cells,sec:unsteady sims}, respectively, and the associated bifurcation diagrams are given in~\cref{fig:codim2 bif 3L,fig:codim2 bif 2L,fig:codim2 bif parity}. Note that the diagrams for $L=2\lambda,3\lambda$ are reproduced from~\cite{hughes2024travelling}. However, some branches are shown with different domain lengths compared to~\cite{hughes2024travelling}, which can change the stability along the branch.

\begin{figure}[!tp]
    \centering
    \includegraphics[width=\linewidth]{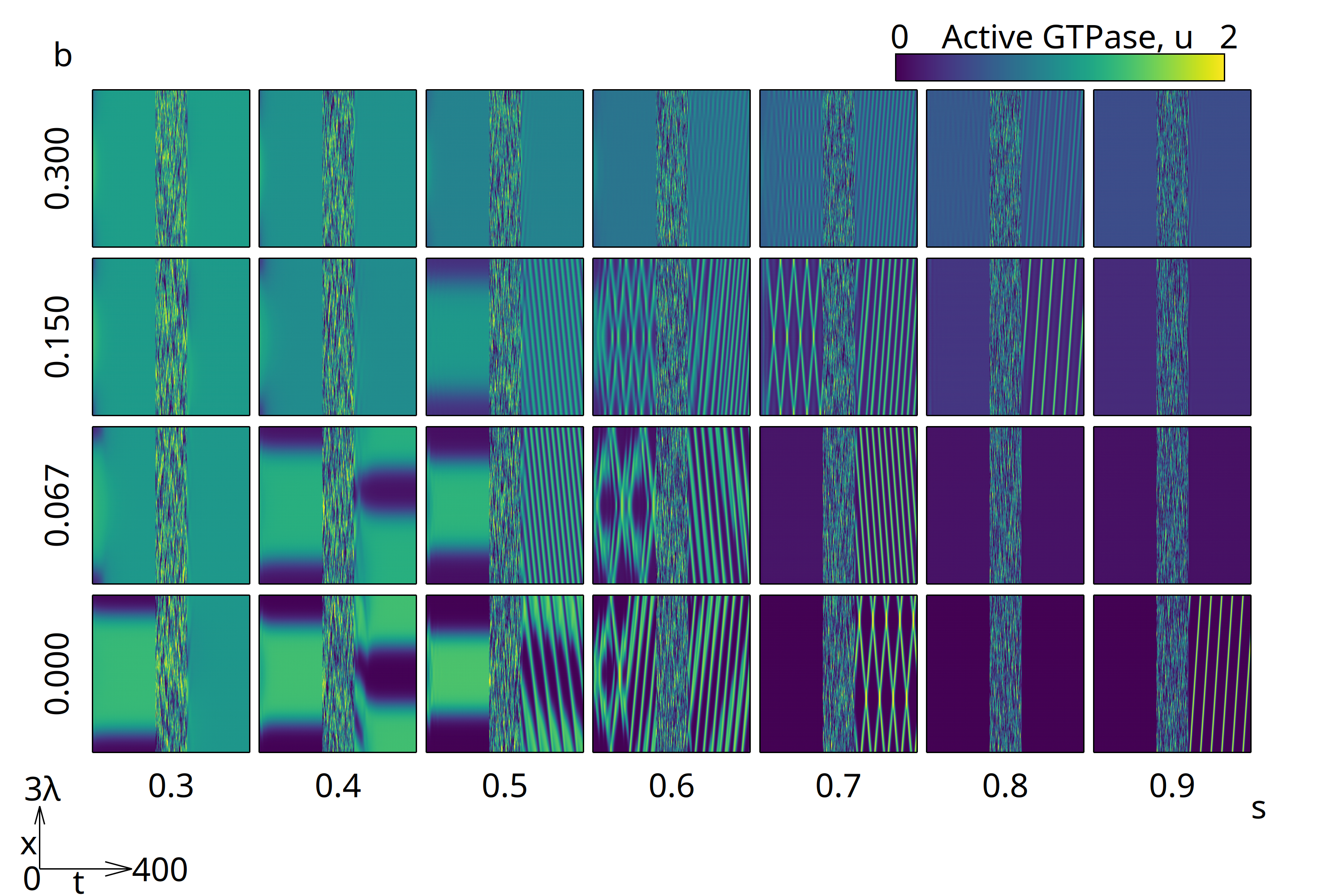}
    \caption{\textbf{Coexistence of polar and ruffling states in time-dependent simulations.} Parameter sweep in the F-actin dependent inactivation rate, $s$, and the GTPase basal activation rate, $b$, showing full PDE solutions with $L=3\lambda\approx9.28$, where $\lambda$ is the wavelength that induces the finite wavenumber Hopf instability of the upper HSS. Solutions are shown as space-time plots of the active GTPase, $u(x,t)$ as a heat map, for several values of $b$ (0, 0.067, 0.15, 0.3 on vertical axis) and various values of $s$ (horizontal axis, $0.3\le s \le 0.9$). The system is initiated with $u(x,0)=0.75-0.5 \cos(2\pi x/L)$, $v(x,0)=1.25-0.1 \cos(2\pi x/L)$, and $F(x,0)=3.5 - 2 \cos(2\pi x/L)$, and simulated to $t=400$. For $160<t<240$, white noise is added to the $du/dt$ equation and subtracted from the $dv/dt$ equation (to avoid changing the total mass $M$). The noise results in several transitions between coexisting states such as polar, TWs, uniform steady states, and more complex time-periodic dynamics. If $s$ is too small or too large, only the uniform state exists. Corresponding profiles of $u, v, F$ at $t=400$ are shown in~\cref{fig:sbProfiles}. Other parameter values as in~\cref{tab:par valuesSims}. }
    \label{fig:sbKymoScan}
\end{figure}

\begin{figure}[!tp]
    \centering
    \includegraphics[width=\linewidth]{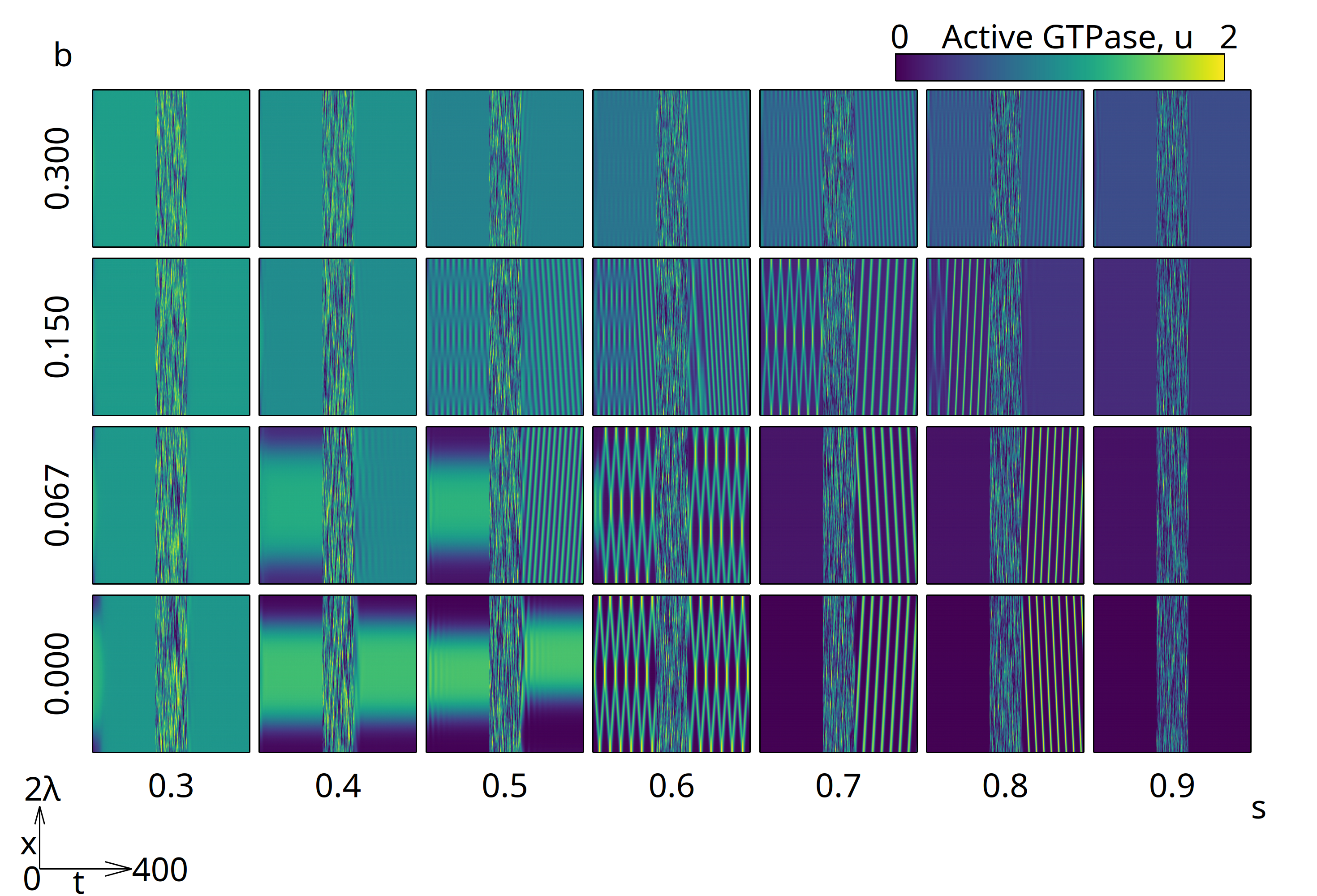}
    \caption{\textbf{Parameter sweep with $L=2\lambda$:} As in~\cref{fig:sbKymoScan} but with $L=2\lambda$. When $L=2\lambda$, time-periodic solutions resembling counter-propagating waves and coexistence of polar and traveling wave solutions are still observed. Other parameter values as in~\cref{tab:par valuesSims}. }
    \label{fig:sbKymoScan 2L}
\end{figure}

\begin{figure}[!tp]
    \centering
    \includegraphics[width=\linewidth]{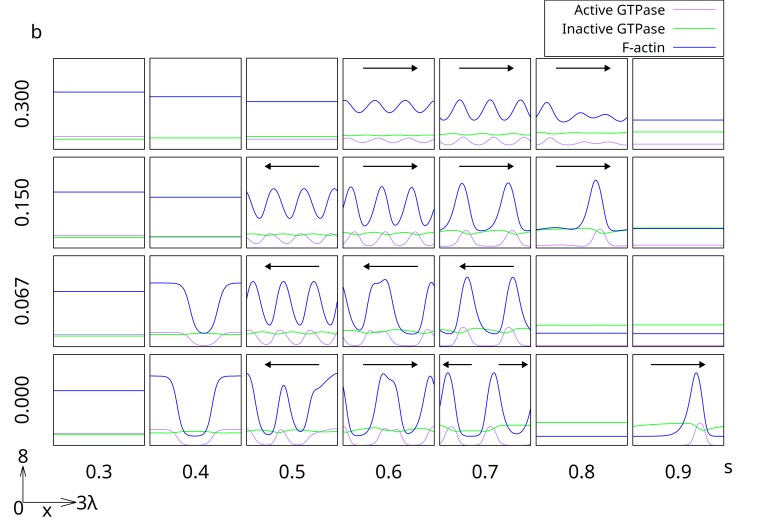}
    \caption{\textbf{Final solution profiles from~\cref{fig:sbKymoScan}.} Line plots showing the $u,v,F$ profiles at $t=400$ for the simulations in~\cref{fig:sbKymoScan}. We see uniform states for small and large values of $s$, traveling waves of 3, 2, or 1 wavelengths for intermediate values of $s$, and polar patterns for $s=0.4$ and small $b$. The inactive GTPase (green) is nearly uniform due to its relatively high rate of diffusion. The arrows indicate the direction of propagation. When $b=0$ and $s=0.7$, counter-propagating peaks are observed and the arrows demonstrate the direction each peak travels.}
    \label{fig:sbProfiles}
\end{figure}

\subsection{Coexistence and transitions between regimes}

To gain intuition for the basic solution structure and to explore distinct regimes, we simulated our model~\cref{eq:model}, with various values of $s$, $b$, and $L$. All other parameters are as in~\cref{tab:par valuesSims}. In order to test for coexisting (bistable) behavior, we introduce random noise for a short time period to scramble the solution and permit a transition to another behavior that exists for the same parameter set. When $L=3\lambda$, several such transitions are evident in~\cref{fig:sbKymoScan}, from polar to wave-like solutions ($s \approx 0.5$ and $0\le b\le 0.15$), and from uniform to wave-like solutions ($0.7\le s\le 0.9 $ and $0\le b\le 0.15$). Note that initially the solutions are symmetric, e.g. polar and uniform patterns, because of the choice of initial condition. We also see cases where waves move in one direction, or in opposite directions. We found that if $s$ is too high or too small, no wave-type solutions were possible. \Cref{fig:sbKymoScan 2L} is another parameter sweep with $L=2\lambda$, showing that fewer types of transient dynamics are possible with smaller domain lengths but the overall long run dynamics are similar. These full PDE solutions confirm the predictions of the previous bifurcation study by \cite{hughes2024travelling}.

We further plotted the profiles of all three variables, $u,v,F$ at the final time point ($t=400$) to demonstrate the relationship of the model components. As shown in~\cref{fig:sbProfiles}, the inactive GTPase is nearly spatially uniform, while the dominant patterns reside in $u,F$. (This follows from the fact that $v$ diffuses fastest, so that it rapidly redistributes along the domain.) Furthermore, we note that for the propagating solutions, such as the traveling waves, the peak of $F$ trails after the peak of $u$ (see the arrows in~\cref{fig:sbProfiles} that denote the direction of propagation). This behavior arises from the diffusion coefficients, that is, $D_F<D_u$.

\subsection{Simulating cell shapes and motility}
\label{sec:sim motility}

We use the cellular Potts model (CPM) to represent fluctuating cell shapes and cell motion.
We chose the open-source multiscale software platform, Morpheus \citep{starruss2014morpheus}, for several reasons: (1) Most of the advanced computational algorithms, including many kinds of PDE simulations are built in \citep{starruss2014morpheus}. (2) Cells are represented as fluctuating stochastic shapes, appropriate to the dynamic shapes of real cells \citep{maree2007cellular}. (3) Morpheus is easy to learn, use, and teach, with basic GUI and visually appealing output images. We have used it extensively as a tool for training novice modelers. (4) Morpheus preserves simulation conditions in small xml files, making every result reproducible (for each of our figures). Our code can be found at the \href{https://gitlab.com/jupiteralgorta/f-actin-on-cell-motility}{GitLab repository}. 

Morpheus has a built-in CPM simulator that enables solutions of PDEs such as~\cref{eq:model} along a cell edge (implemented as a \emph{membrane property})\footnote{Technically, the PDEs are solved on the rim of a circle with the same area as the CPM cell, then mapped onto the CPM cell edge.}. Those PDE solutions are linked to protrusion and retraction of that edge via Morpheus plugins (See~\cref{app:sim details} for details). Specifically, locations along the cell with high F-actin concentration are more likely to protrude outward and then the area constraint of the CPM promotes retraction. 

The cellular Potts model is a lattice-based simulation where a single cell is represented by a connected subset of lattice sites (pixels) with an ID=1. Sites outside the cell (denoted \emph{medium}) are labeled with ID=0. Cell expansion and retraction can occur along the edge of this cell. In its simplest form, the algorithm proceeds as follows:
\begin{enumerate}
    \item Choose a random lattice site $i$.
    \item Choose a random neighboring lattice site $j$ and propose to copy the ID from $i$ to $j$.
    \item Calculate the difference in the free energy (\emph{Hamiltonian}), $\Delta H$, between the current and proposed new configuration, where the Hamiltonian, $H$, is given by
    \[
    H=\lambda_a (A-A_0)^2 + \lambda_p (P-P_0)^2,
    \]
    with $A, A_0, \lambda_a$ being the current area, target area, and area constraint weight, and similarly for the perimeter $P$. 
    \item Accept or reject the new configuration according to the following rule:
    \begin{itemize}
        \item If $\Delta H\leq0$, accept;
        \item If $\Delta H>0$, then accept the new configuration with a probability $e^{-\Delta H/T}$, where $T$ is the Boltzmann temperature parameter.
    \end{itemize}
\end{enumerate}

While the canonical CPM selects sites randomly from the entire lattice, Morpheus \citep{starruss2014morpheus}, the software package we use, optimizes this by selecting only sites at the boundaries between cell and medium where configuration changes can occur. In our implementation, we also apply the perimeter constraint using an aspherity parameter, which instead computes the perimeter relative to that of a circle with the same area as the CPM cell. The area constraint implies that cell protrusion along part of the edge is inherently balanced by a tendency to retract other parts of the cell and the perimeter constraint acts as an effective membrane tension or stiffness that governs overall cell shape.

To capture the effects of F-actin on cell shape and motility, the PDEs~\cref{eq:model} are solved at each point on the CPM cell edge. Updates to the CPM simulation that protrude the cell edge at $x$, a location along the cell edge, are favored if the value of F-actin, $F(x,t)$, is high. This results in an additional term added to the Hamiltonian. For each simulation, the cell is initiated as a circle. 
More details about the CPM simulations, parameter values, and other technical matters are given in the~\cref{app:sim details}.

\section{Polar cells and ruffling states ($L=3\lambda$)}
\label{sec:CPMResults}

As mentioned, we will now connect the PDE solutions to a moving cell framework via CPM simulations that depict cell shapes and migration. We set basic parameters at their default values (\cref{tab:par valuesSims}), fix $b=b_c\approx0.067$ and $L=3\lambda$, and vary the parameter $s$. In each case, we pick values of $s$ and initial conditions close to a polar or traveling wave solution, and display PDE solutions as space-time plots of $u(x,t)$. In all simulations there is some initial transients as the system is not exactly at a steady state. In some cases, the same color scheme also shows $u(x,t_i)$, at various times $t_i$, along the edge of the 2D CPM cell. We also show a time sequence of cell shapes and locations to demonstrate dynamics and motion of those cells. The parameter values and initial conditions used in the simulation are described in~\cref{app:sim details}, and preserved in the xml files in the GitLab repository. 

\subsection{Bifurcation structure}
\begin{figure}[!tp]
    \centering
    \includegraphics[width=\textwidth]{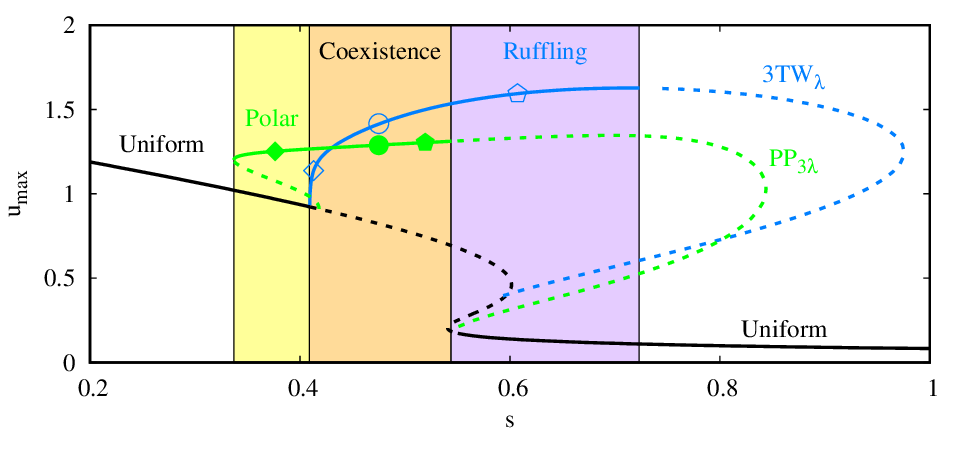}
    \caption{\textbf{Coexistence of polar and ruffling states:} 
    Single parameter bifurcation diagram with respect to the negative feedback parameter, $s$, for $b=b_c\approx0.067$ and $L=3\lambda$, where $\lambda\approx3.09$ is the wavelength that leads to the instability of the upper HSS. The solutions are projected onto the maximum value of the active GTPase, $u_{\max}$. Solid lines denote linearly stable solutions (i.e., observable in the long run) and dashed lines denote unstable solutions. Uniform HSSs (solid black curves) represent  unpolarized resting cells with spatially uniform GTPase activity around its perimeter. Polarized states with wavelength $3\lambda$ (PP$_{3\lambda}$) are stable along the green solid curve. Traveling waves with three peaks and wavelength $\lambda$ (3TW$_\lambda$, associated with \emph{ruffling}) are stable along the solid blue curve. 
    As $s$ increases, transitions in behavior are predicted: from uniform to polar, from polar to ruffling, and from ruffling to low uniform stable state. However, not all possible solution types are illustrated. For example, traveling waves with 1 and 2 peaks and bipolar and tripolar patterns also exist in this regime (see \cite{hughes2024travelling} and~\cref{fig:codim2 bif 3L full} for more details). In the orange shaded region, both polar and ruffling states coexist, so transitions between them can take place.
    Typical solutions are shown in the $b=b_c\approx0.067$ row of~\cref{fig:sbKymoScan}. The shapes and motility behavior of cells corresponding to points along these curves are shown in~\cref{fig:Polar_Solutions,fig:WobblyCell,fig:3TW_Solutions}.} \label{fig:codim2 bif 3L}
\end{figure}

From the simulations of the model~\cref{eq:model} in~\cref{fig:sbKymoScan}, we see that certain parameter settings admit more than one possible behavior, that is, there is coexistence of distinct solutions types. Transitions between those types were elicited by noise, without changing parameter values. We wish to understand the regimes of coexistence by considering the one-parameter bifurcation diagrams of the PDEs. The results, reproduced from~\cite{hughes2024travelling}, are summarized in~\cref{fig:codim2 bif 3L}. Here, as $s$ is varied, a codimension-2 long wavelength/finite wavenumber Hopf bifurcation is observed (although not fully captured as $L$ is small; see \cite{hughes2024travelling} for more details). The non-uniform solutions include 3-peak traveling waves with wavelength $\lambda$ (3TW$_\lambda$; \emph{ruffling}) that coexist with uni-polar patterns of wavelength $3\lambda$ (PP$_{3\lambda}$; \emph{polarization}).

The bifurcation diagram in~\cref{fig:codim2 bif 3L} informs us of several trends: (1) Varying $s$ leads to transitions between stable states: from a uniform state with no pattern (low s, white region), to a polar state (yellow shaded region), to a wave-like ruffling state (purple), and back to uniform. These transitions are depicted by the endpoints of solid curves in each case. It is important to note that other solution types exist in this regime, including bipolar and tripolar patterns, traveling waves with 1 and 2 peaks (see \cite{hughes2024travelling} and~\cref{fig:codim2 bif 3L full} for more details), and multi-frequency oscillatory states or other solutions bearing some resemblance to polar or wave-like states. The bipolar and tripolar patterns are all unstable (a characteristic property of mass-conserved systems~\citep{weyer2023mesa}) and so we focus on unipolar patterns. In particular, we 
focus on
the solutions shown in~\cref{fig:codim2 bif 3L}. (2) There is a substantial range of values of $s$ for which both polar and ruffling states coexist and are stable (overlap of yellow and purple regions, shown as orange region). (3) Other states (dashed lines) exist in various parameter regimes, but are unstable, and would not be observed biologically in the long term (but could lead to transient behavior). (4) Polar cell states exist for lower negative feedback strengths $s$ whereas traveling waves exist for higher $s$ values. (5) The structure of the solutions and the bifurcation points (at which stability changes occur or new solutions emerge) can be classified in terms of known dynamical systems transitions, shedding light on their generic (model independent) properties. For example, mass conservation is necessary for the polar patterns to exist for broad parameter regimes. In particular, mass conservation allows for such patterns to form without coexistence of HSSs and causes them to not propagate in space~\citep{verschueren2017model}.

\subsection{Polar cells display directed motion}
\begin{figure}[!tp]
    \centering
    \includegraphics[width=\linewidth]{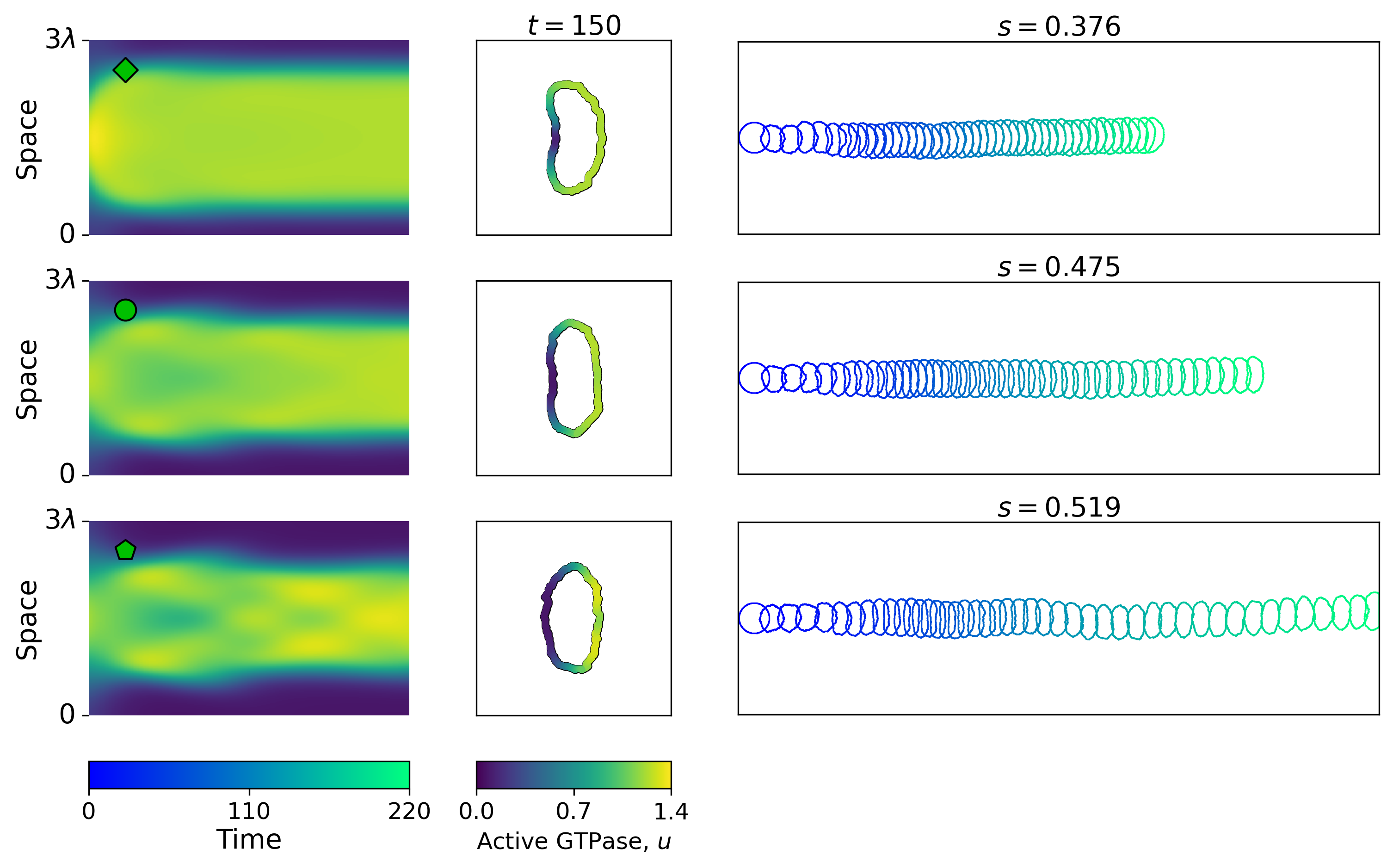}
    \caption{\textbf{Shapes and trajectories of polar cells corresponding to points on the polar branch in~\cref{fig:codim2 bif 3L}:}
    Three simulations of PDE solutions in space-time plots (left), resulting cell shapes (center) and motion (right) simulated in Morpheus \citep{starruss2014morpheus}. The cell locations over time are shown (right) with heatmap for $0\le t \le 220$. The results correspond to the green symbols on the (polar) branch in~\cref{fig:codim2 bif 3L}, for $s=0.376, 0.475, 0.519$, i.e., the cell perimeter $L=3\lambda$. Some transients are shown in each case. Note that as $s$ increases, the width of the Rho zone shrinks, making Rho more highly concentrated at the front. This results in increased $F$ (not shown) and greater cell speed. It also accounts for a transition from canoe-shaped to D-shaped keratocyte-like cells \citep{keren2008mechanism} after some transients.
    Initial conditions provided in~\cref{eq:ICfig_Polar_Solutions}, $b=b_c\approx0.067$, and other parameter values as in~\cref{tab:par valuesSims}. Morpheus file: \href{https://gitlab.com/jupiteralgorta/f-actin-on-cell-motility/-/tree/main/Morpheus_code/Polar?ref_type=heads}{Polar Files}.}
    \label{fig:Polar_Solutions}
\end{figure}

We first consider the dynamics of cells corresponding to points marked by green shapes along the green polarity branch in~\cref{fig:codim2 bif 3L}. Results are shown in~\cref{fig:Polar_Solutions} for values of $s=0.376, 0.475, 0.519$ (green diamond, circle, and pentagon, respectively, in~\cref{fig:codim2 bif 3L}). The initial conditions lead to some transients while the distribution of components along the cell edge settles into its steady polar profile. During these transients, the cell shape oscillates but directed cell motion persists. 
 
The fastest, and most polarized cell is found at $s=0.519$. This cell has the narrowest, tallest plateau of $u$, seen from the width of the yellow band in the space-time plots. This value of $s$ leads to the most focused cell front, where the active GTPase and F-actin are densest. We can understand the cell speed from the fact that protrusion is assumed to be proportional to $F$, and $F$ tends to a value proportional to $u$ in the model. Hence, a higher GTPase level at the front implies a higher F-actin density and stronger protrusion force.

The steady state shape of the cell also depends on $s$. This is demonstrated in the central column of~\cref{fig:Polar_Solutions} where both steady state cell outlines and values of $u$ along the edge are shown. As $s$ increases, the cell morphology changes from canoe-shaped to D-shaped \citep{keren2008mechanism}. On the whole, these simulations bear a significant resemblance to the polarized motion of keratocytes.

\subsection{Ruffling cells do not migrate}

\begin{figure}[!tp]
    \centering
    \includegraphics[width=\linewidth]{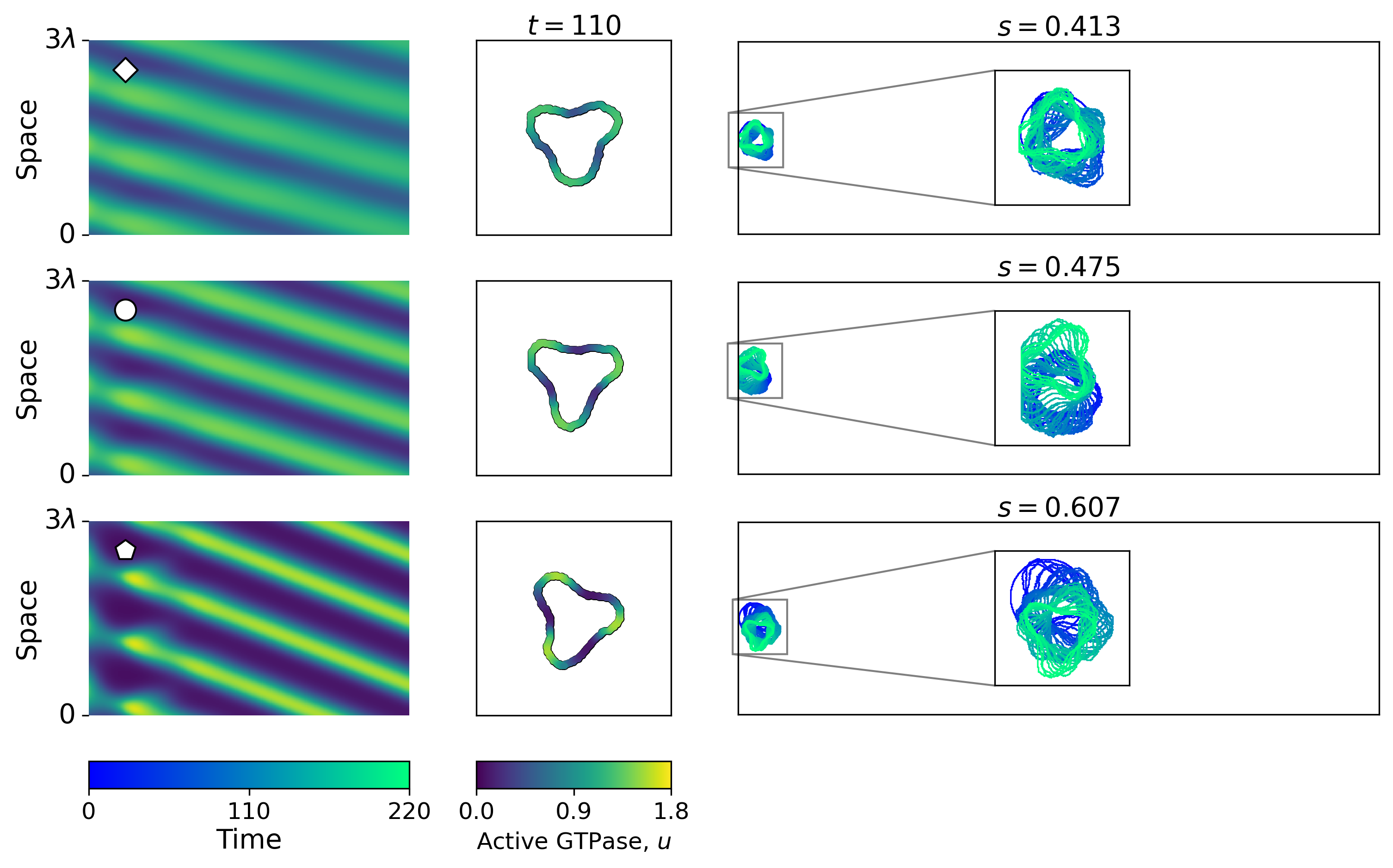}
    \caption{\textbf{A \emph{ruffling} (3TW) cell hardly moves}: As in~\cref{fig:Polar_Solutions}, but for the traveling wave branch in~\cref{fig:codim2 bif 3L} with $s=0.413, 0.475, 0.607$.  Parameter values corresponds to shapes along the blue branch in~\cref{fig:codim2 bif 3L}. The insets of the right panels show a zoomed-in view of the various cell shapes. Note that the cells have three competing protrusions, and have no directed migration. This figure was produced with the same parameter values as in~\cref{fig:Polar_Solutions} (\cref{tab:par valuesSims}), but with different values of $s$ and different initial conditions (see~\cref{eq:IC3TW}). Morpheus files: \href{https://gitlab.com/jupiteralgorta/f-actin-on-cell-motility/-/tree/main/Morpheus_code/TW?ref_type=heads}{3TW Files}.}
    \label{fig:3TW_Solutions}
\end{figure}

We initiated cells with the same set of basic parameters (other than $s$), but with a distinct initial distribution of components along their edge, to demonstrate the behavior close to the ruffling branch in~\cref{fig:codim2 bif 3L}. In particular, we used initial conditions with three peaks that have an initial asymmetry in the model components so that propagation is favored. Results are shown in~\cref{fig:3TW_Solutions}.

After some transients, the PDE solutions form traveling waves with three peaks. Correspondingly, each of the simulated cells settle into a shape with three protrusions circulating around the cell perimeter. As a result, there is no unique cell front. Directed migration is lost entirely, and the cells appear to rotate in situ.
The rotation speed increases with larger $s$ values, due to
higher propagation speed of the traveling wave in the underlying PDE system~\cref{eq:model}. Note that the middle rows of~\cref{fig:Polar_Solutions,fig:3TW_Solutions} have identical parameter values, and differ only in initial conditions. This verifies the stable coexistence of polar and ruffling cell states.

The waves of these ruffling model cells are reminiscent of waves of edge protrusion seen in the Sheetz lab \citep{giannone2004periodic} in mouse embryonic fibroblasts (MEFs) (see their SI movies 5, 7, 14), and was later quantified 
in MEFs, fly wing disk cells, and mouse T cells (Figure 2 in \cite{dobereiner2006lateral}). Typical edge protrusion speeds observed there were 0.5-5 $\mu$m/min.

\subsection{Destabilized polar cells have dynamic shapes}
\begin{figure}[!tp]
    \centering
    \includegraphics[width=0.95\linewidth]{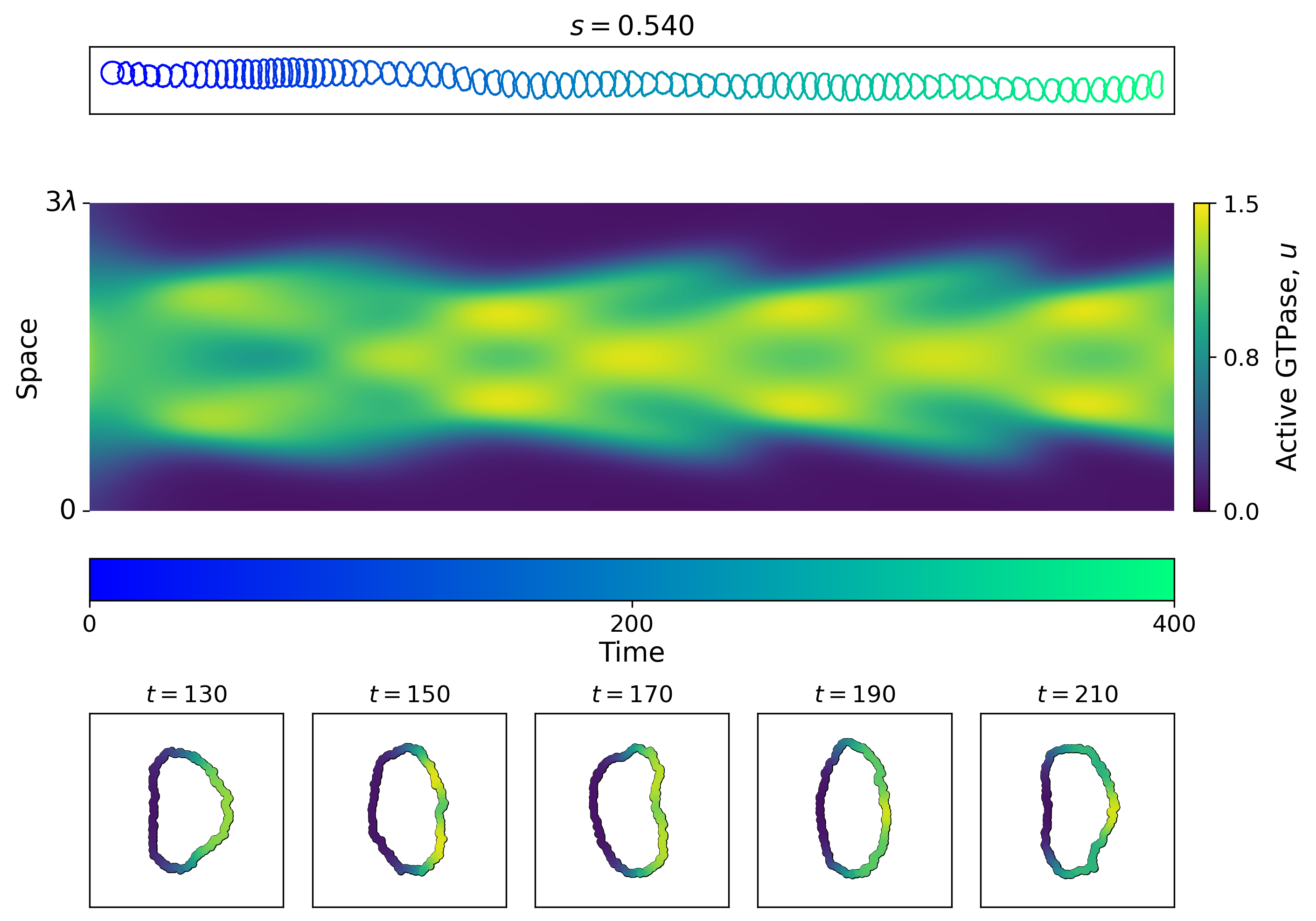}
    \caption{\textbf{Destabilized polar cells:} The shapes and motion of a cell that is initiated with the same conditions as~\cref{fig:Polar_Solutions}, but with $s = 0.54$, i.e., after the instability of polar patterns. Here, the cell locations over time are shown at the top, the PDE solution as a space-time plot in the middle, and active GTPase concentration along the cell edge at the bottom. Note the oscillations in cell shape and speed. Produced with \href{https://gitlab.com/jupiteralgorta/f-actin-on-cell-motility/-/blob/main/Morpheus_code/Wobby_s_0.600.xml?ref_type=heads}{Wobby\_s\_0.600.xml}.}
    \label{fig:WobblyCell}
\end{figure}

For the given parameters (see~\cref{tab:par valuesSims}) and cell perimeter $L=3\lambda$, polarization destabilizes at $s\approx0.54$. From \cite{hughes2024travelling}, we know that the instability is a Hopf bifurcation, suggesting the emergence of oscillations around the polar distribution. In \cite{hughes2025chaos}, a simulation governed by this instability was investigated in the context of real cells hitting a wall. In particular, Hughes~\textit{et al} demonstrated that the underlying instability captured the initial transition from directed to disordered motion post collision. Here, we show how cell shape and migration are affected by this instability in the context of CPM simulations. \Cref{fig:WobblyCell} shows the resulting simulation. We used the same initial conditions as in~\cref{fig:Polar_Solutions}, but with $s=0.55$, which is just past the onset of instability of the polar patterns.

After some initial transients, the distribution of GTPase settles into a time-periodic oscillation around a polar distribution. The distribution of the GTPase still maintains a region of high activity that does not translate along the edge of the cell. This causes the cell to maintain directed cell motion. However, the cell  oscillates between a D-shape and a canoe  (or kidney) shape. The peak of the GTPase is not always in the direction of motion. In the canoe shapes, there are two competing cell-fronts that coordinate to maintain directed motion. The simulation in~\cref{fig:WobblyCell} shows that at this instability, directed cell motion persists. Therefore, to fully transition to disordered motion, additional instabilities, stochastic effects, higher dimensionality, or more complex signaling networks are needed (as predicted in \cite{hughes2025chaos}).

\section{Exotic trajectories and coexisting modes in smaller cells ($L=2\lambda$)} \label{sec:smaller cells}

\subsection{Bifurcation structure}

\begin{figure}[!tp]
    \centering
    \includegraphics[width=\textwidth]{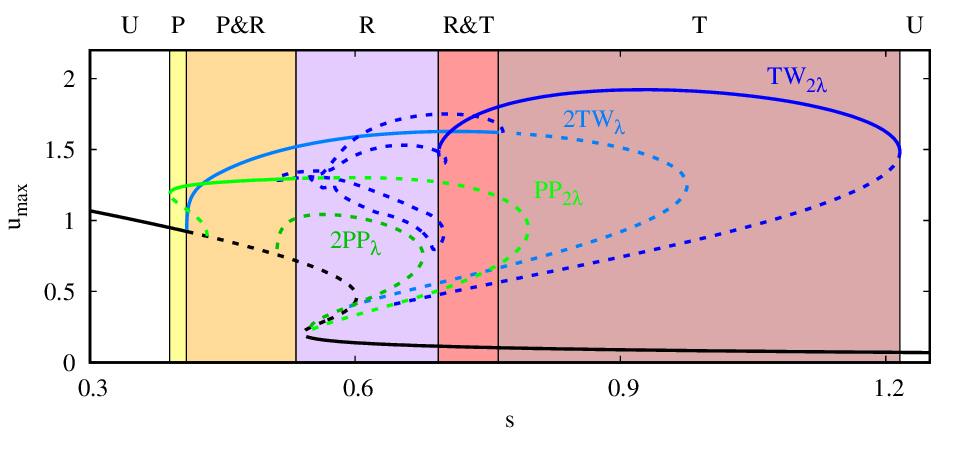}
    \caption{\textbf{Coexistence of polar and ruffling states in smaller cells:} 
    Single parameter bifurcation diagram as in~\cref{fig:codim2 bif 3L} but with $L=2\lambda$. This figure shows part of the bifurcation structure for simulations in~\cref{fig:CW2_solution,fig:CW_Solution}. Here, we show unipolar (PP$_{2\lambda}$) and bipolar (2PP$_{\lambda}$) solutions, and traveling waves with 1 peak (TW$_{2\lambda}$, a turning cell) and 2 peaks (2TW$_\lambda$, a ruffling cell). The shaded regions denote various existence and coexistence regimes of nonlinear patterns: yellow - polarization (P), orange - polarization and ruffling (P\&R), purple - ruffling (R), red - ruffling and turning (R\&T), and brown - turning (T). Cell turning with a perimeter $L=2\lambda$ is shown in the top panel of~\cref{fig:CW_Solution}.} \label{fig:codim2 bif 2L}
\end{figure}

From the simulations in~\cref{fig:sbKymoScan,fig:sbKymoScan 2L}, we see that time-periodic patterns can form, resembling counter-propagating peaks. In the following simulations we use $L=2\lambda$, $s\in\{0.6,0.8\}$ and retain all other parameters. We choose $L=2\lambda$ because at $L=3\lambda$, the resulting solutions have additional peaks that interfere with the counter-propagating peaks. The bifurcation diagram for cells of this size is shown in~\cref{fig:codim2 bif 2L}. However, the time-periodic solutions that are observed are not presented in that figure. Note that the distinct cell size produces distinct ranges of stability of polar and wave-like solutions but the bifurcation structure is similar to~\cref{fig:codim2 bif 3L}, i.e., coexistence of uni-polar and multi-peak traveling waves. Here, we explore the results for counter-propagating peaks as well as single traveling waves. 

\subsection{Counter-moving protrusions}

\begin{figure}[!tp]
    \centering
    \includegraphics[width=\linewidth]{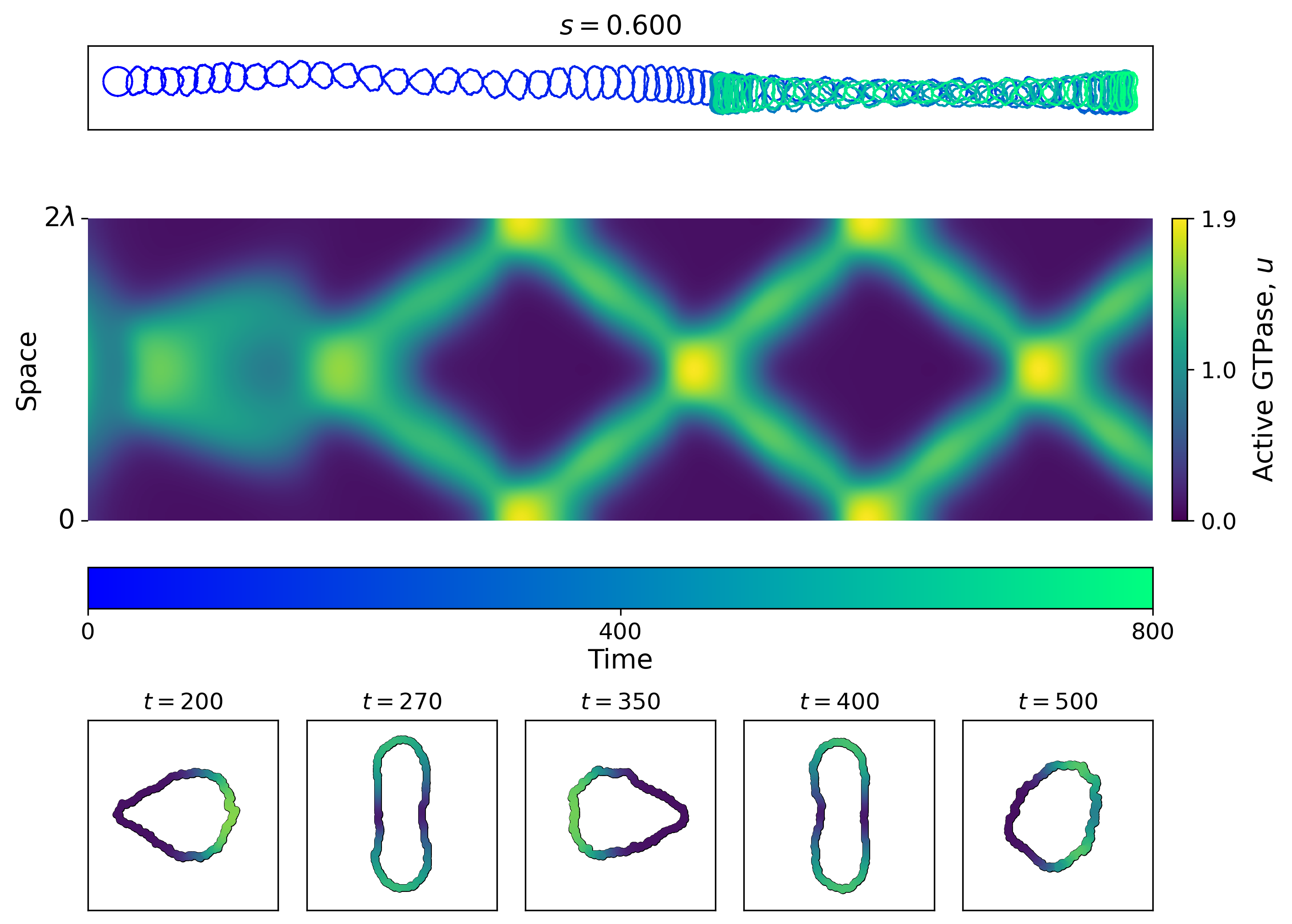}
    \caption{\textbf{Coexisting motility modes:} 
    A simulation as in~\cref{fig:WobblyCell} but in a smaller cell (with perimeter $L=2\lambda$, where $\lambda$ is the critical wavelength of the finite wavenumber Hopf instability) and with $s=0.6$. The cell, initiated in a polar state, starts to oscillate. Two counter-propagating peaks then get established, which oscillate back and forth. The cell moves forward and back as the waves propagate around its rim. Initial conditions given in~\cref{eq:ICsCW2_Solutions}. The cell continues in its wobbly periodically shifting state. Morpheus file: \href{https://gitlab.com/jupiteralgorta/f-actin-on-cell-motility/-/blob/main/Morpheus_code/CW_s_0.600.xml?ref_type=heads}{CW\_s\_0.600.xml}. }
    \label{fig:CW2_solution}
\end{figure}

We ran simulations with cells initialized with polar initial conditions but with $s=0.6$, which is after the instability onset of polar patterns (PP$_{2\lambda}$). As shown in~\cref{fig:CW2_solution}, the polarized distribution rapidly destabilized, first into a few standing oscillations (left portion of space-time plot), and then into a pair of traveling waves (TW) moving in opposite directions.  These counter-moving peaks result in two protrusions circulating around the cell's edge. 

While the two protrusions are close to one another, they coordinate and determine an effective cell-front propelling the cell in a compromise direction.  As the waves reach the cell rear, the cell reverses its direction of motion. As seen in~\cref{fig:CW2_solution}, this sets up a back and forth migration pattern. The shapes of the cell also fluctuate periodically, with two protrusions merging and splitting along the cell edge. This cycle of shapes and motion stems from our simplified cell-edge domain, and would not be likely to occur in real cells, where waves are prevented from reaching the cell rear.

\subsection{Cells with one or two protrusions}    
\begin{figure}
    \centering
\includegraphics[width=\linewidth]{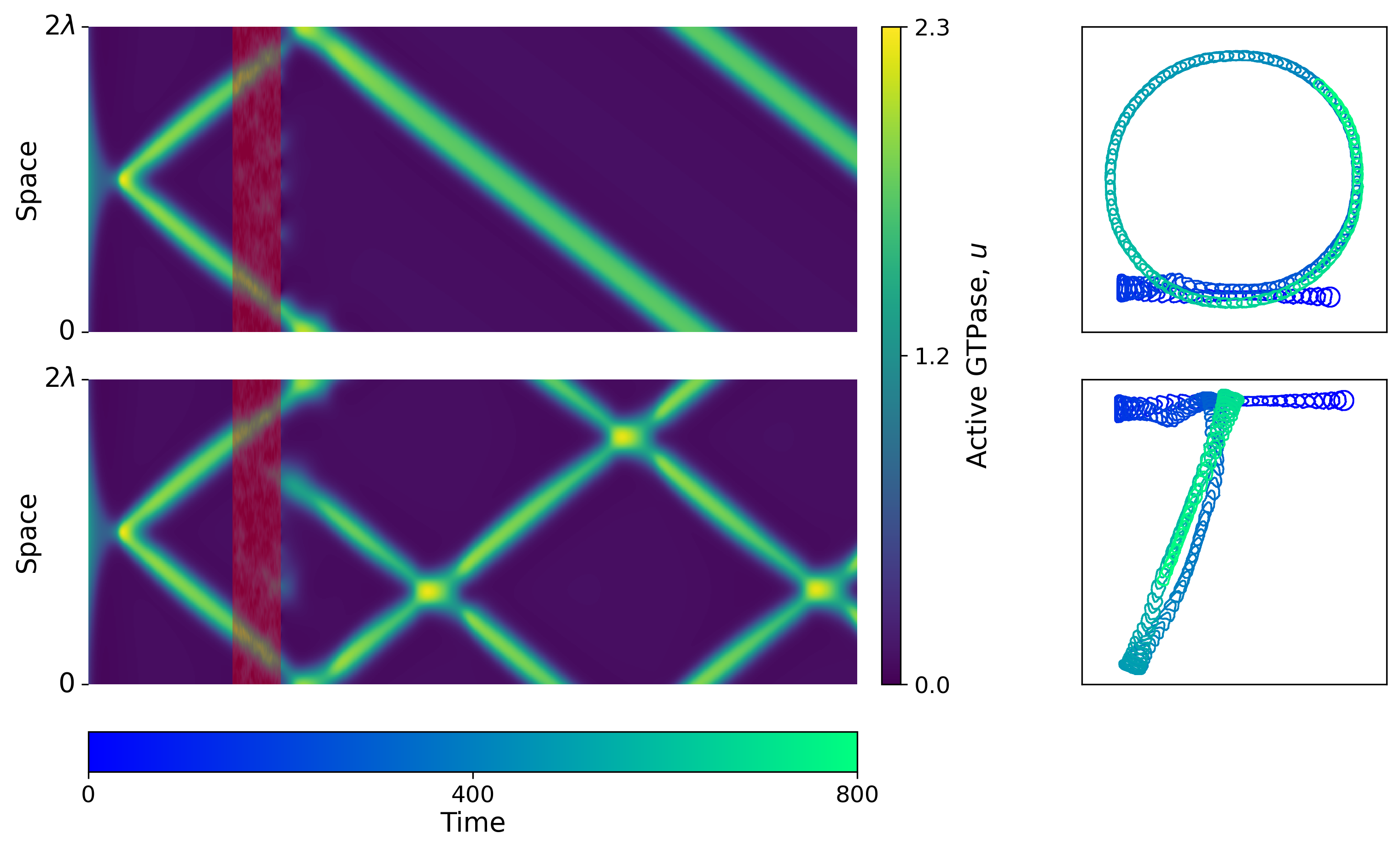}
    \caption{\textbf{Exotic trajectories:} As in~\cref{fig:CW2_solution}, but for $s=0.8$ and with noise at $150<t<200$ (red stripe). The resulting single TWs or counter-propagating peaks lead to small circular trajectories (top) or reversal of cell motion (bottom). Initial conditions and other parameters as in~\cref{fig:CW2_solution}. Morpheus file: \href{https://gitlab.com/jupiteralgorta/f-actin-on-cell-motility/-/blob/main/Morpheus_code/CW_s_0.8_noise.xml?ref_type=heads}{CW\_s\_0.8\_noise.xml}. }
    \label{fig:CW_Solution}
\end{figure}

For larger $s$, ruffling states destabilize, and single-peak traveling waves and counter-propagating peaks persist.
To visualize transitions between these states, we initiated simulations in the same way as in~\cref{fig:CW2_solution}, and added noise after some time, to see transitions. Results, shown in~\cref{fig:CW_Solution}, demonstrate the distinct cell behavior when TWs or a pair of counter-moving peaks are present.

In both cases, counter-propagating peaks establish early in the simulation. Following application of noise (red band on space-time plot), roughly 9 out of 10 cases reverted to the same counter-propagating peaks. In a few cases, including top panel of~\cref{fig:CW_Solution}, only one peak persisted after the noise.

When a single-peak TW remains, the cell has a unique front that circulates around its edge. Consequently, the cell trajectory is circular, with a polarize skewed cell shape. This contrasts with the counter-propagating peaks case, where the cell moves back and forth, as previously found. In the latter case, the noise changes the axis of migration, but not the overall behavior. These results confirm the coexistence of dynamical states such as single peak TWs and time-periodic solutions in the model.

Single and multiple traveling waves have been observed in real cells. For example, \cite{machacek2006morphodynamic} found that mouse embryonic fibroblasts (MEFs) and PtK1 epithelial (canoe) cells display modulated traveling waves moving bidirectionally around the cell edge (e.g., \cite{machacek2006morphodynamic} Fig 10). For PtK1 cells, the authors found noisy counter-propagating traveling waves with wave-speed of $\approx6\mu$m/min. 

\subsection{Static resting cells at low and high values of $s$}

We explored several other settings with similar parameters but for large and small values of $s$ that are outside of the pattern-forming range (results not shown). For low and high values of $s$, e.g., $s=0.2$ and $s=1.3$, respectively, all nonuniform patterns shown cease to exist, and only a uniform spatial distribution of $u,v,F$ is stable. In these cases, an initially polar cell rapidly loses polarity and stalls. The distance that an initially polarized cell can move in that case depends on the timescale of the $u, v, F$ components relative to the persistence time assumed in the cell simulations.

\section{Simulations with spatial-temporal parameters} \label{sec:unsteady sims}

The cell-scale simulations presented thus far are all performed with constant parameter values. These results demonstrate likely long-run behavior but provide little detail on transitions between different solution types. In particular, neither the bifurcation diagram nor the simulations demonstrate how to transition between directed motion and turning, nor how continual spatio-temporal noise affects these results. We focus on these questions here.

\subsection{Transitions for time-dependent F-actin negative feedback.} \label{sec:time s}

\begin{figure}[!tp]
    \centering
    \includegraphics[width=\textwidth]{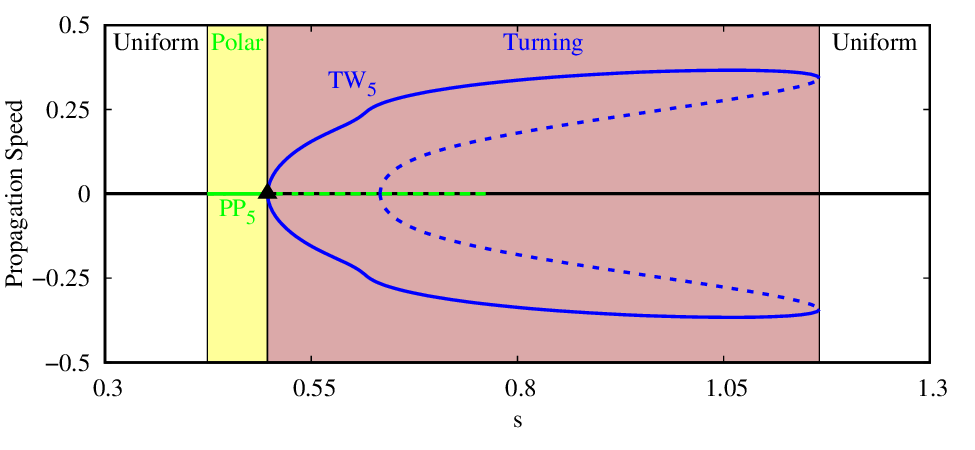}
    \caption{\textbf{Transition between polarization and turning:} 
    Single parameter bifurcation diagram as in~\cref{fig:codim2 bif 3L} but with $L=5$ and projected onto the propagation speed of the solution. In this diagram the traveling wave solutions (TW$_5$) have 1 peak and represent cell turning. Here, we see that there is no coexistence between static polar patterns (PP$_5$) and traveling waves. Instead, the polar patterns lose stability at a parity-breaking bifurcation (at $s\approx0.497$, black triangle) caused by the loss of reflective symmetry (i.e., loss of the invariant $x\to-x$, $u\to u$), which leads to the emergence of stable traveling waves. Note that as in~\cref{fig:codim2 bif 3L}, not all possible solutions are shown. Simulations investigating the transition from polar to cell turning as $s$ is varied are shown in~\cref{fig:POLAR2TW,fig:TW2POLAR,fig:Random_motion}.
    } \label{fig:codim2 bif parity}
\end{figure}

For $L=3\lambda/2$, \cite{hughes2024travelling} observed a narrow parameter regime with continuous transition between polarization and cell turning as $s$ is varied. Here, setting $L=5$ for a wider regime, we demonstrate the behavior (see~\cref{fig:codim2 bif parity}). While no coexistence is observed at this parameter setting, we do find a continuous transition between polarization and cell turning. In particular, the stable traveling waves emerge from a so-called parity-breaking bifurcation of the stable polar solutions (at $s\approx0.50$), where separate branches of left and rightward moving traveling waves emerge with the same wavelength. This structure suggests a possible mechanism for transitioning between cell turning and polarity by slowly varying $s$ between the two regimes of stability.

\begin{figure}
    \centering
    \includegraphics[width=\linewidth]{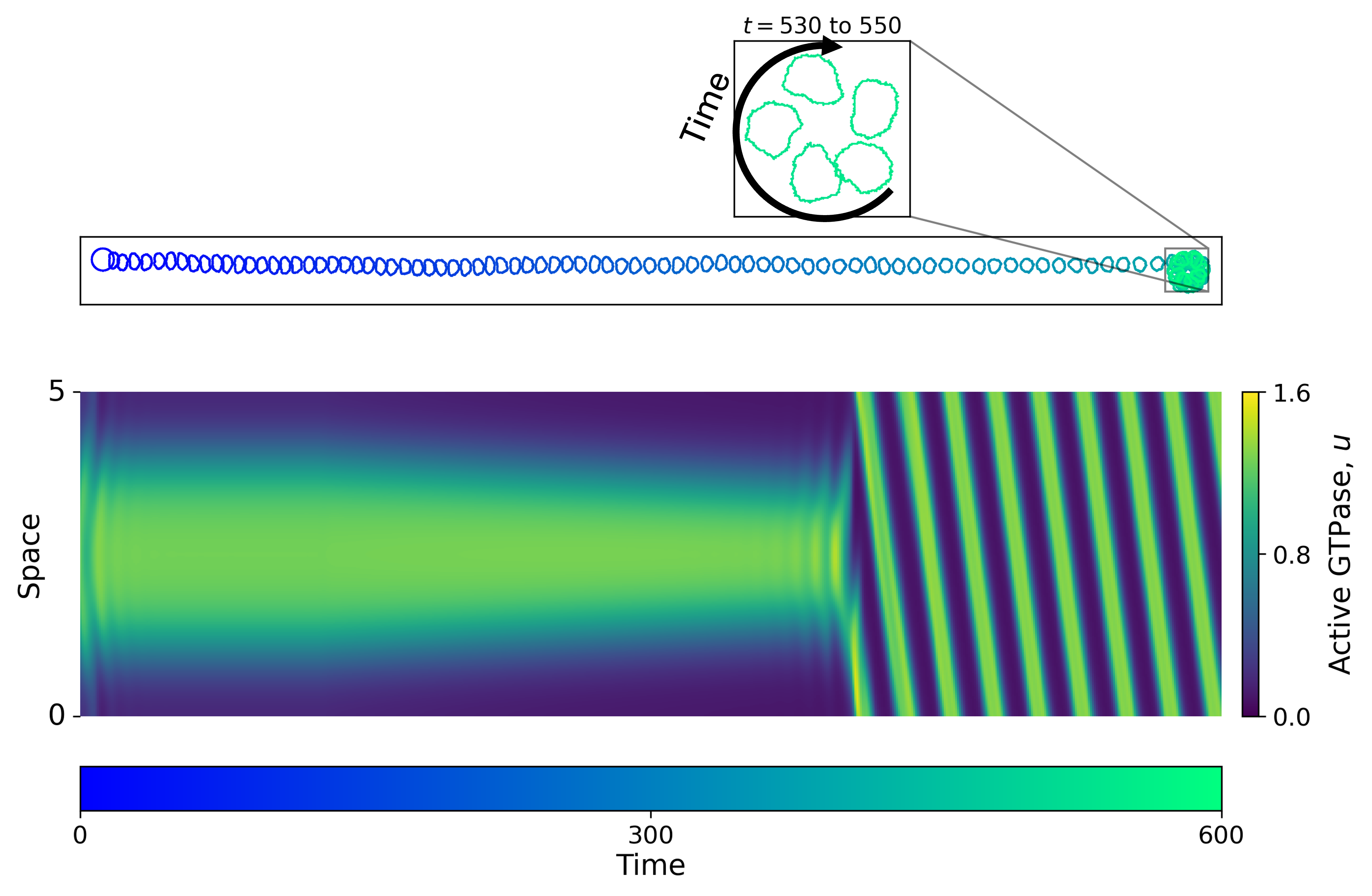}
    \caption{\textbf{Transition from directed cell motion to cell turning:} The simulation starts with a polar initial state~\cref{eq:ICsCW2_Solutions}, and $s = 0.475$. At time $t= 125$, the value of $s$ grows linearly according to~\cref{eq:Increas_s}, attaining $s = 0.7$ at $t= 375$. This causes the polar pattern to become a single traveling wave. Space-time plot of $u$ (bottom) and cell shapes and trajectory (top) are shown. The inset shows one full cycle of the cell shapes along the circular path induced by the single-peak traveling wave. Parameter values as in~\cref{tab:par valuesSims} with $L=5$, $b=b_c\approx0.067$, and $s=s(t)$ given by~\cref{eq:Increas_s} with $s_0=0.475$ and $s_f=0.6$. Produced with Morpheus file \href{https://gitlab.com/jupiteralgorta/f-actin-on-cell-motility/-/blob/main/Morpheus_code/POLAR2TW.xml?ref_type=heads}{POLAR2TW.xml}}
    \label{fig:POLAR2TW}
\end{figure}

\begin{figure}
    \centering
    \includegraphics[width=\linewidth]{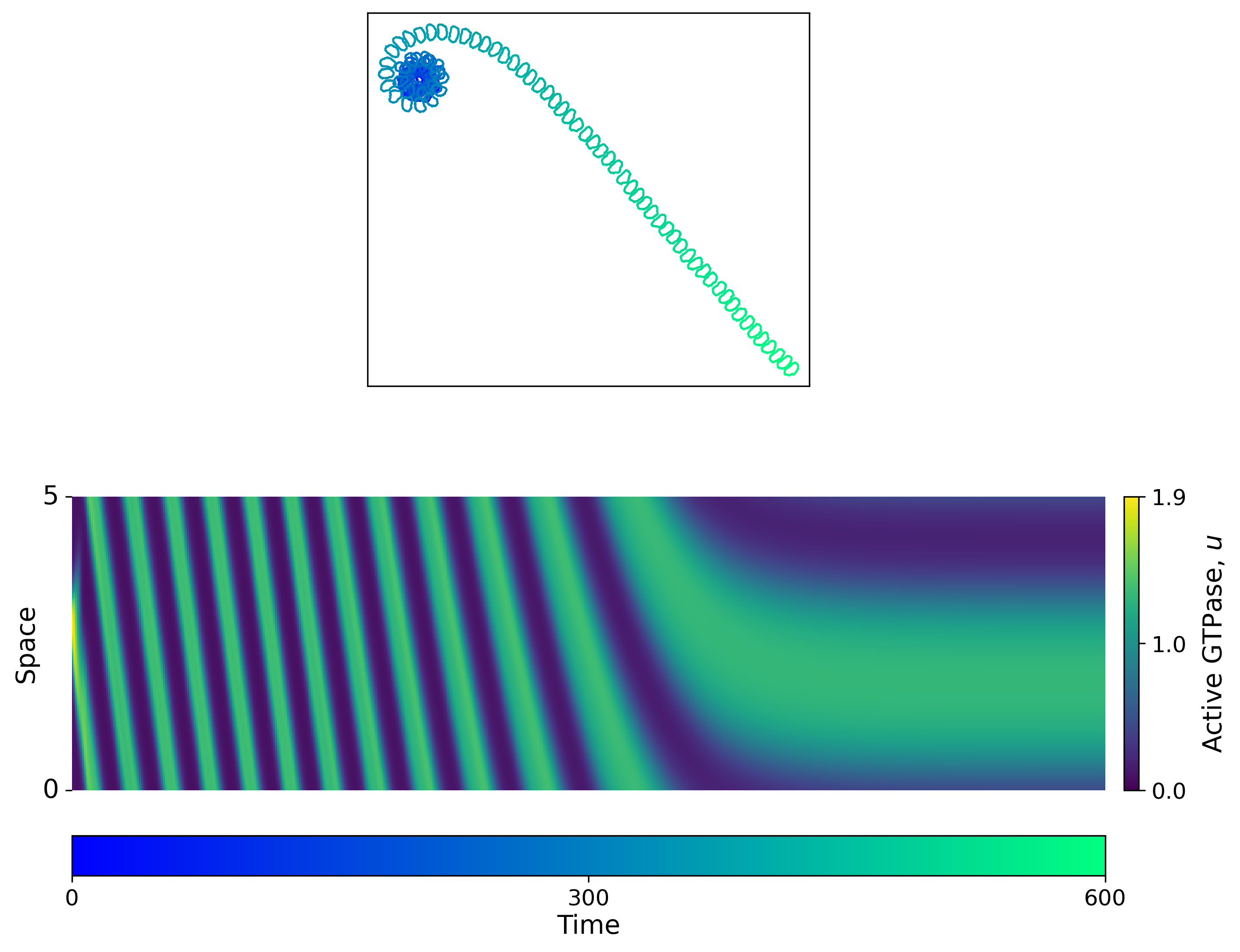}
    \caption{\textbf{Transition from cell turning to directed cell motion:} 
    As in~\cref{fig:POLAR2TW}, but with $s_0=0.6$ (for $0 \le t \le 125$) and $s_f=0.475$ (for $t \ge 375$). Initial conditions given by~\cref{eq:ICsTW2WP_Solutions} form a traveling wave that turns into a polar solution as $s$ decreases. The cell starts by traversing a circular path as in the end of the simulation shown in~\cref{fig:POLAR2TW} (blue part of image, top), and later moves in a polarized manner towards the bottom-right. Produced with Morpheus file \href{https://gitlab.com/jupiteralgorta/f-actin-on-cell-motility/-/blob/main/Morpheus_code/TW2POLAR.xml?ref_type=heads}{TW2POLAR.xml}.}
    \label{fig:TW2POLAR}
\end{figure}

\begin{figure}[!tp]
    \centering
    \includegraphics[width=\linewidth]{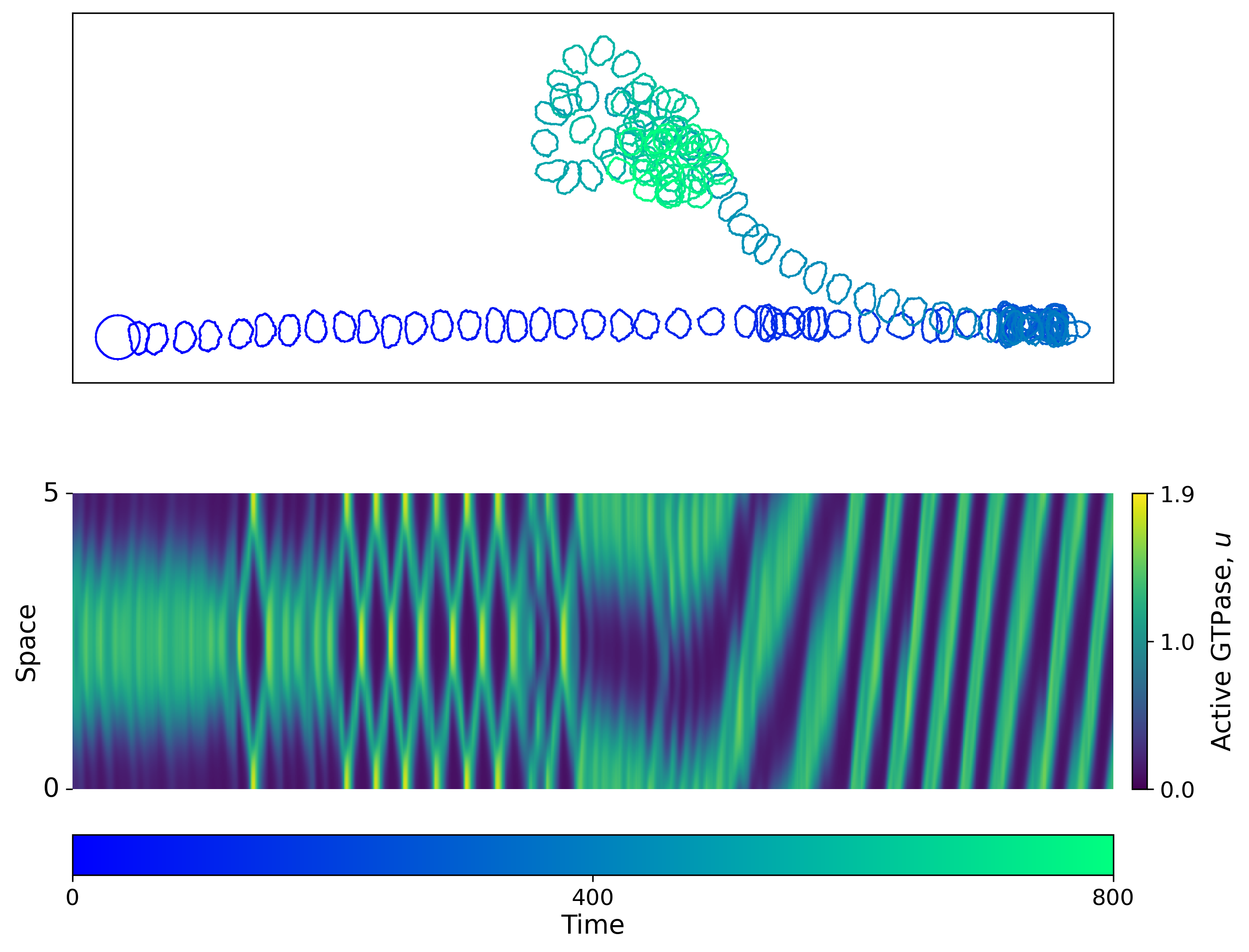}
    \caption{\textbf{Randomly changing $s(t)$:}
    As in~\cref{fig:POLAR2TW,fig:TW2POLAR}, but with $s(t)$ constantly evolving randomly within the interval $0.475\le s \le 0.6$ based on a Weiner process (essentially, Brownian motion with reflective boundaries at $s = 0.475$ and $s = 0.6$.) We see random switching between polar, counter-propagating peaks, and traveling waves. Each run is unique, but all start with the polar configuration. Produced with Morpheus file \href{https://gitlab.com/jupiteralgorta/f-actin-on-cell-motility/-/blob/main/Morpheus_code/Random_motion.xml?ref_type=heads}{Random\_motion.xml}.}
    \label{fig:Random_motion}
\end{figure}

Motivated by~\cref{fig:codim2 bif parity}, we provide simulations with a time varying F-actin dependent inactivation rate, $s(t)$. In particular, we vary $s$ so that we cross this parity-breaking bifurcation and see transitions between polar cells and cells with a single traveling wave along its perimeter. The domain length is set to $L=5$ in these simulations to match that of~\cref{fig:codim2 bif parity}. Results are shown in~\cref{fig:POLAR2TW,fig:TW2POLAR,fig:Random_motion}.

We first consider a linear time-dependent increasing ramp for $s(t)$ over the range $0.475\le s\le 0.7$.
This ramp sweeps across the polar and TW regimes of the model. As shown in~\cref{fig:POLAR2TW} (left to right), the cell is initially polar, and moves directionally rightwards. As $s$ increases, the width of the plateau region of high GTPase decreases, and is eventually replaced by a single-copy traveling wave. As a result, over the final 200 time units, the cell traverses a circular path showing cell turning (see the inset of~\cref{fig:POLAR2TW} showing one revolution of the circular path). Notably these transitions are relatively sharp, as predicted by our bifurcation analysis. 

A similar transition, but for a decreasing ramp of $s(t)$ over the range $0.6 \ge s \ge 0.475$ is shown in~\cref{fig:TW2POLAR}. As $s$ decreases, the propagation speed of the traveling wave decreases to zero, and then a polar distribution emerges. Afterwards, the cell moves along an oblique path to the bottom-right of the domain.

Finally, we allow $s(t)$ to vary randomly within the range $0.475\le s \le 0.6$. (This was implemented as a kind of random walk of the parameter $s$ within this range, with reflective boundaries at the interval endpoints.) As shown in~\cref{fig:Random_motion}, the resulting space-time plots demonstrate transitions in behavior. The cell shapes and trajectory (left to right) produce a mixture of the motility modes discussed above, and the cell moves randomly in its domain. Interestingly, for $100\leq t\leq 400$, we also see counter-moving peaks as in~\cref{fig:CW_Solution,fig:CW2_solution}. This suggests that $s$ is varying too fast and the polar patterns destabilize into these periodic solutions as seen in the aforementioned figures.

\subsection{Continual spatio-temporal noise}

\begin{figure}[!tp]
    \centering  
    \includegraphics[width=\linewidth]{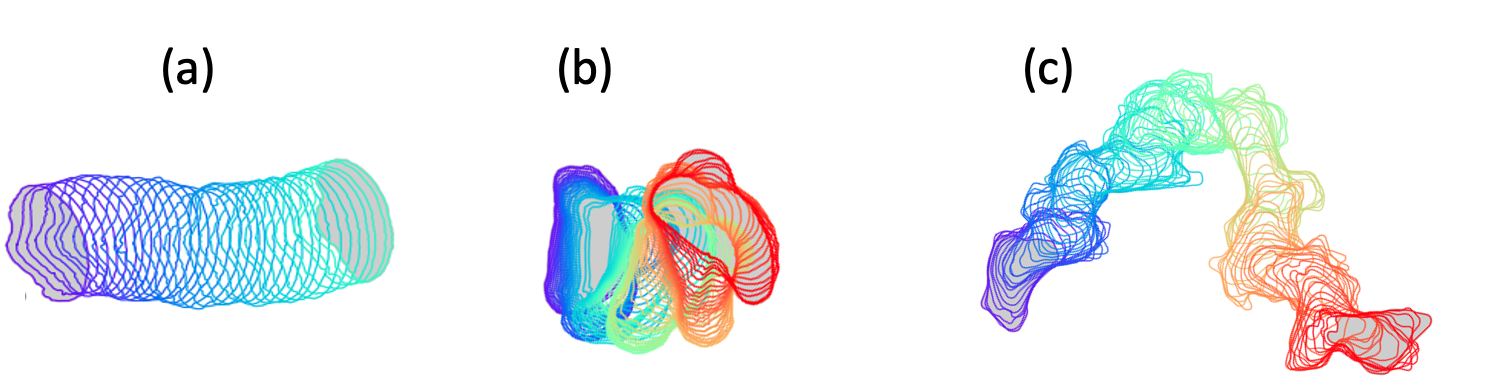}
    \caption{\textbf{Distinct modes of cell motility in real cells from \cite{schindler2021analysis}:}
    Cell contours (time increases from purple to light green and then red) for (a,b) polar and turning keratocytes and (c) amoeboid motion of a \emph{Dictyostelium discoideum} cell. Adapted from Figs 6 and 11 in \cite{schindler2021analysis}, published under the \href{https://creativecommons.org/licenses/by/4.0/}{CC BY 4.0 License}, in PLOS One.}
    \label{fig:SchindlerShapes}
\end{figure}

The minimal model for actin regulation~\cref{eq:model}, together with the basic CPM simulation would not be expected to capture the complexity of real cells. Nevertheless, several examples described have aspects that resemble motion observed in real cells. This is true of the polarized motion of keratocytes, shown in the experimental cell contour data from \cite{schindler2021analysis}. Two samples of this type, for polar and turning keratocytes are reproduced in~\cref{fig:SchindlerShapes}a,b. 

However, cells of the social amoeba \emph{Dictyostelium discoideum} display less regular migration, with alternations between dominance of one or another pseudopod \citep{insall2010understanding,neilson2011modeling,hughes2025chaos}. This gives rise to the kind of dynamics shown in~\cref{fig:SchindlerShapes}c. 
We asked whether our prototypical model could capture any aspects of such behavior given the addition of spatially-dependent periodic noise that interferes with one or the other waves moving around the cell.

\begin{figure}[!tp]
    \centering
    \includegraphics[width=\linewidth]{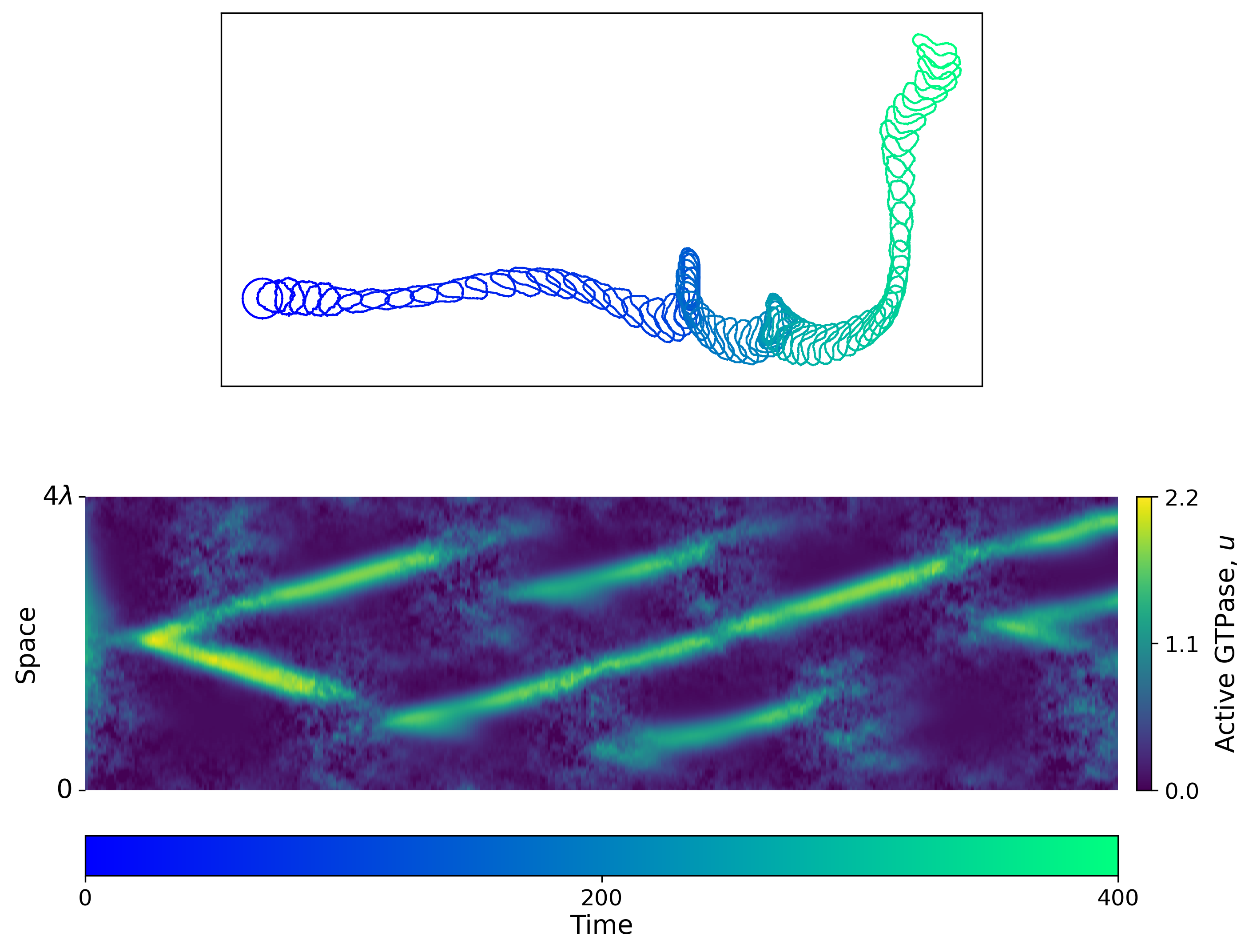}
    \caption{\textbf{Effect of continual noise.} A simulation, as in~\cref{fig:CW_Solution}, but with continual space-time distributed noise~\cref{eq:Dicty noise} that disrupts the counter-propagating peaks and $L=4\lambda$. Random noise can cause one or the other pseudopod to take over, in a behavior resembling the Dictyostelium motility phenotype. Parameters as in~\cref{tab:par valuesSims} with $s=0.8$. Initial conditions given by~\cref{eq:DictyType}.
    Produced with \href{https://gitlab.com/jupiteralgorta/f-actin-on-cell-motility/-/blob/main/Morpheus_code/Dicty.xml?ref_type=heads}{Dicty.xml}}
    \label{fig:DictyType}
\end{figure}

To probe this possibility, we initiated a cell with the same conditions as in~\cref{fig:CW_Solution}, but with $L=4\lambda$ and noise at one or the opposite cell edge with periodic intensity. (See~\cref{app:sim details} for details.) (We chose a larger domain as more complex dynamics can emerge.) Results of this trial are shown in~\cref{fig:DictyType}. The space-time plot demonstrates that protrusions (bright stripes of activity of $u$) alternate moving right versus left in an irregular manner. This affects both the cell shapes and cell trajectory. In this basic model variant, we cannot claim to adequately account for the Dictyostelium motion. However, we do see that pseudopod dominance and cell trajectories have some features in common with~\cref{fig:SchindlerShapes}c. For more detailed pseudopod-competition model of Dictyostelium, see \cite{neilson2011modeling}.

\section{Discussion} 

Our main goal in this paper was to leverage recent analysis of a prototypical model for F-actin regulation to reveal its implications for cell shapes and migration.
The existing theory consisted of numerical PDE bifurcation analysis of a system of active and inactive GTPase, and F-actin \citep{hughes2024travelling}. This basis helped us to find interesting regimes of parameters and understand how transitions in behavior depend on parameters. Among the parameters of interest, we chiefly investigated the F-actin dependent GTPase inactivation rate. The bifurcation analysis also allowed us to construct relevant simulation experiments and interpret the results.

The model~\cref{eq:model} for the regulatory circuit described in~\cref{fig:SchematicJHBio}b is closely related to models for actin waves proposed and studied earlier \citep{holmes2012,Liu2021}. A few 2D CPM cell simulations  were shown in \cite{Liu2021}, but here, we conducted a systematic study of how the PDE bifurcations lead to cell behavioral transitions. To do so, we simplified the representation of cells by retaining only their edge contours, and solved the reaction-diffusion PDEs on the cell edge using the Morpheus open-source software plugin for a membrane RD solver. Further, these simulations linked edge protrusion at each site to the magnitude of the F-actin variable $F$ at that site. Retraction was implicitly computed from an area constraint in the CPM.

The simulations demonstrate a variety of behaviors, such as directed motion of polar and wobbly cells, in situ ruffling of cells whose edge sustains traveling waves, cell turning, and pseudopod competition. The addition of stochasticity can also produce the less regular amoeboid behavior observed in \emph{D. discoideum}.
Of particular interest is the observation that a fixed set of parameters can result in coexisting motility modes (e.g. polar or ruffling cells), so that noise or input of some kind can cause transitions in cell behavior.

Our simulations and bifurcation analyses show that these different phenotypes and regimes of behavior are governed by key model parameters, and thus we can make predictions of the types of observed cell shapes and motility based on variation of these parameters. In particular, we observed that the width of the polar pattern, and thus the width of the leading edge, is inversely proportional to the F-actin-dependent inactivation rate, $s$. We also saw that lower values of $s$ favor polarity and higher values favor turning and ruffling. The cell perimeter, $L$, or equivalently the diffusion rates of the GTPase and F-actin or the GTPase basal rate of inactivation (see~\cref{sec:signal model}) also play an important role in the observed phenotypes. The transition between stable polarity and single-peak traveling waves, through the parity-breaking bifurcation, is only observed for a specific range of $L$. For $L$ below this range, single-peak waves are observed but stable polarity is absent.

The simplicity of the model and its simulations also mean that predictions have limitations. First, the membrane RD solvers in Morpheus do not take cell edge deformation into account: rather, the RD system is solved on the edge of a circular cell of equal area and mapped onto the deforming cell edge. Second, we do not control explicitly for membrane tension; the CPM perimeter constraint acts to limit edge expansion, but does not affect the signaling system.
Finally, we assume that parameters are constant along the cell edge, whereas in real cells, the front and back likely become distinct, and prevent waves from circulating all around the cell. See, for example, lamellipodial waves in \cite{barnhart2017adhesion} that propagate only along the front part of the cell edge.

All in all, our paper emphasizes several key ideas: (1) Mathematical analysis of prototypical models for signaling can guide judicious exploration of model parameters and facilitate numerical exploration. (2) PDE bifurcation analysis can provide insights into coexistence of behaviors and how new behaviors emerge as parameters are tuned.
(3) Relatively simple CPM simulations can provide a useful tool for exploratory simulations of the link between regulatory circuits and emergent cell behaviors.

Elsewhere \citep{algorta2025investigating}, a similar computational approach was recently used to link proposed Rac signaling circuits to optogenetic experimental data for migrating neutrophils. Despite its limitations, the CPM simulation platform could readily capture realistic cell trajectories for many stimuli protocols. In future, we plan to investigate models for interacting GTPases (e.g., Rac-Rho), effects of membrane tension, and nonuniform environments to extend the realism of the modeling approach.

\backmatter

\section*{Statements and Declarations}

\bmhead{Conflicts of Interest} The authors have no conflicts of interest to declare relevant to this article’s content.

\bmhead{Data availability:} The code files for the Morpheus simulations are linked in the associated figures.

\appendix

\section{Parameter values at the codimension-2 instability} \label{app:codim2}
Using default parameter values from \cref{tab:par valuesSims},
\cite{hughes2024travelling} varied $s$ and $b$, finding the codimension-2 long wavelength/finite wavenumber Hopf instability at $s=s_c\approx0.41$ and $b=b_c=0.067$. The finite wavenumber Hopf instability has an intrinsic wavelength of $L=\lambda\approx3.09$ and the long wavelength occurs for $L\gg\lambda$.

\section{More complete bifurcation diagram with $L=3\lambda$} \label{app:more bifs}
\begin{figure}[!tp]
    \centering
    \includegraphics[width=\textwidth]{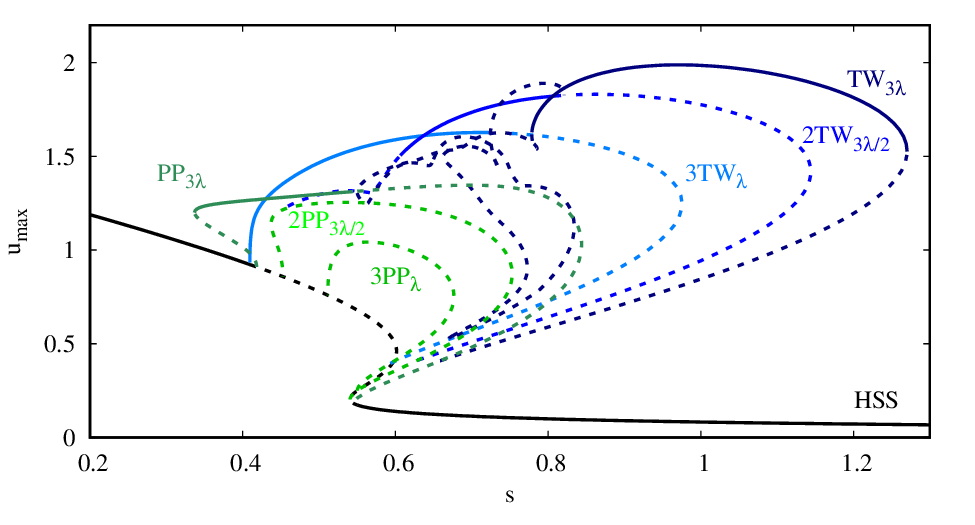}
    \caption{\textbf{Polar and traveling waves of various wavelength:} As in~\cref{fig:codim2 bif 3L} but showcasing traveling waves (TWs) and polar patterns (PPs) with wavelengths $\lambda$, $3\lambda/2$ and $3\lambda$. The notation 3TW$_{\lambda}$ denotes TWs with wavelength $\lambda$ and three spatial copies. This figure shows part of the bifurcation structure for simulations in~\cref{fig:Polar_Solutions,fig:WobblyCell,fig:3TW_Solutions}. The traveling wave solutions represent cell turning (TW$_{3\lambda}$) and ruffling (2TW$_{3\lambda/2}$ and 3TW$_\lambda$). The polar solutions PP$_{3\lambda}$, 2PP$_{3\lambda/2}$, and 3PP$_\lambda$ represent polar, bipolar, and tripolar solutions, respectively. However, as before, only stable solutions (solid branches) are biologically observable.} \label{fig:codim2 bif 3L full}
\end{figure}

\Cref{fig:codim2 bif 3L full} shows all stationary solutions emerging from the HSSs when $L=3\lambda$ and the traveling wave solutions with the same wavelength. We include this figure to demonstrate that traveling waves with different wavelengths also exist/coexist and only unipolar patterns are stable. 

\section{Simulation details} \label{app:sim details}

All simulation codes are available for each of the results in our figures. Here we briefly summarize simulation conditions.
We use a rectangular lattice, typically $600\times 600$ pixels, with periodic boundary conditions. In cases where polarized cells travel for longer distances, the lattice is extended up to $3000\times 600$ to see the entire trajectory. Simulations are typically run for 400-800 MCS, as indicated in the figure timescales. Our CPM cell has target area $A_0=4800$ and area constraint strength $\lambda_a=1$. Rather than fixing a target perimeter, we constrain the perimeter toward $P = \phi\,2\sqrt{\pi A}$, where $A$ is the current cell area and $\phi=1.2$ is the aspherity ($\phi = 1$ represents a circle). This constraint has a strength $\lambda_p=1$. (Both these constraints are built-in Morpheus options.) We use the Metropolis Kinetics with yield=0.05 and temperature $T=1$. The cell is initiated as a circle of area $A_0$. PDEs are solved using the built-in Runge-Kutta O(4) method, and tested extensively against simulations of the same system (on a static 1D domain) in Julia. We use the Morpheus plugin \emph{SurfaceMotion} that links the local concentration of F-actin to a bias that favors local protrusion of the cell edge. This is implemented in Morpheus by assigning a Hamiltonian term of the form:
\[
H_{\mathrm{F}} = \lambda_{\mathrm{F}}\, F(x,t),
\]
where $F(x,t)$ is the F-actin variable in the PDEs. A pixel on the cell edge extending into the surrounding medium (protrusion)  will be accepted with probability:
\[
P(\Delta H, H_{\mathrm{F}}) = 
  \begin{cases}
  1, & \text{if } (\Delta H - H_{\mathrm{F}})\le 0,\\
  \exp \left(- \frac{\Delta H - H_{\mathrm{F}}}{T}\right), & \text{if } (\Delta H - H_{\mathrm{F}}) > 0.
  \end{cases}
\]

Furthermore, to represent a kind of persistence of cell motion due to gradual cytoskeletal turnover, we used the Morpheus \emph{Protrusion} plugin. This plugin implements the \emph{Act} model \citep{niculescu2015crawling}: newly added pixels are assigned a fixed  ``initial actin level'' (ACT) that decays over time, creating a memory of recent protrusive activity. The CPM Hamiltonian is biased to favor copy attempts from high-activity pixels to neighboring low-activity pixels, mimicking actin-driven edge protrusion. These plugins are built into Morpheus, and users need only set parameter levels as desired.

Unless otherwise stated, we used parameter values given in~\cref{tab:par valuesSims} with $b=b_c\approx0.067$.

\textbf{\Cref{fig:Polar_Solutions}:} 
Domain length: $L=3\lambda$, where $\lambda\approx3.09$ is the numerically solved for wavelength that leads to the finite wavenumber Hopf of the codimension-2 long wavelength/finite wavenumber Hopf. The three simulations use the $s$ values $0.376,0.475,0.519$.

Initial conditions:
\begin{subequations}
\label{eq:ICfig_Polar_Solutions}
\begin{align}
 u(x,0)&=0.75-0.5\cos(x),\\
v(x,0)&=1.25+0.1\cos(x),\\
F(x,0)&= 3.5-2\cos(x).
\end{align} 
\end{subequations}

\bigskip
\textbf{\Cref{fig:3TW_Solutions}:}
Domain length: $L=3\lambda$ and $s$ values $0.413,0.475,0.607$.

Initial conditions:
\begin{subequations}
\label{eq:IC3TW}
\begin{align}
 u(x,0)&=1-0.5\cos(3x)-0.32\sin(3x),\\
v(x,0)&=1+0.028\cos(3x)+0.08\sin(3x),\\
F(x,0)&= 4.6-2\cos(3x).
\end{align} 
\end{subequations}

\bigskip
\textbf{\Cref{fig:WobblyCell}:} Same initial conditions, domain length, and parameters as for~\cref{fig:Polar_Solutions}, but with $s=0.54$.


\bigskip
\textbf{\Cref{fig:CW2_solution}}:
Domain length: $L=2\lambda$ and $s=0.6$.

Initial conditions:
\begin{subequations}
\label{eq:ICsCW2_Solutions}
\begin{align}
 u(x,0)&=0.75-0.5\cos(x),\\
v(x,0)&=1.25+0.1\cos(x),\\
F(x,0)&= 3.5-2\cos(x).
\end{align} 
\end{subequations}
\bigskip

\textbf{\Cref{fig:CW_Solution}:} Same initial conditions, domain length, and parameter values as for~\cref{fig:CW2_solution} but with $s=0.8$ and noise added to the GTPase equations for $150\leq t\leq 200$. In particular, we added 
Noise:
\[
\xi(x,t)=4\mathcal{N},
\]
where $\mathcal{N}$ is a standard normal random number at each grid and time point. The noise was implemented as follows:
\begin{subequations} \label{eq:noise implementation}
    \begin{align}
        \frac{du}{dt}&= (b+\gamma u^2)v - (1+s F+u^2) u+ D\triangle u+\xi(x,t),\\ 
        \frac{dv}{dt}&= -(b+\gamma u^2)v + (1+s F+u^2) u+ \triangle v-\xi(x,t).
    \end{align}
\end{subequations}
to maintain mass conservation.

\bigskip

\textbf{\Cref{fig:POLAR2TW}:}
Domain length: $L=5$.

Parameter values as in~\cref{tab:par valuesSims}, but with a time-dependent F-actin dependent inactivation rate, with parameters $s_0=0.475$ and $s_f=0.6$
\begin{align}
\label{eq:Increas_s}
    s(t)=\begin{cases}
        s_0 & t < 125,\\
        s_0-(s_0-s_f)(t-125)/250 & 125\leq t\leq 375,\\
        s_f & t > 375.
    \end{cases}
\end{align}

Initial conditions given by~\cref{eq:ICsCW2_Solutions}.

\bigskip
\textbf{\Cref{fig:TW2POLAR}:} Same domain length and parameters as for~\cref{fig:POLAR2TW} but with the starting and ending values of $s$ interchanged, i.e., $s_0=0.6$ and $s_f=0.475$. 

Initial conditions: 
\begin{subequations}
\label{eq:ICsTW2WP_Solutions}
\begin{align}
 u(x,0)&=0.07+2\sech(3(x-L/2)),\\
v(x,0)&=M-u(x,0),\\
F(x,0)&= 1.15+5\sech(3(x-L/3)),
\end{align} 
\end{subequations}
where $M=2$.
\bigskip

\textbf{\Cref{fig:Random_motion}:} Same domain length, parameters, and initial conditions as for~\cref{fig:POLAR2TW} but with $s$ changing randomly in time within the interval $[0.475,0.6]$ (see the xml file for implementation details \href{https://1drv.ms/u/c/c293a7caf9ce932c/IQDWmlWvMAIOQ6NrAcEUPP6xAbBduOjAFwfYxVfuIsxRKHs?e=66geor}{Dicty.xml}).
\bigskip

\textbf{\Cref{fig:DictyType}}:
Domain length: $L=4\lambda$ and $s=0.8$. 

Initial conditions:
\begin{subequations}
\label{eq:DictyType}
\begin{align}
 u(x,0)&=1-0.5\cos(x),\\
v(x,0)&=1+0.1\cos(x),\\
F(x,0)&= 4.5-0.82\cos(x).
\end{align} 
\end{subequations}

Continual spatiotemporal noise was added similarly to~\cref{fig:CW_Solution} but of the form
\begin{align} \label{eq:Dicty noise}
\xi(x,t)=2[(1+\sin(x))((1-\cos(6.5 t/100))^2)+(1-\sin(x))(1+\cos(6.5t/100))^2]\mathcal{N},
\end{align}
where $\mathcal{N}$ is a standard normal random number at each grid and time point. The noise was then added and subtracted to the active and inactive GTPase equations as in~\cref{eq:noise implementation}.


\bibliography{Main}

\end{document}